\documentclass[sigconf,10pt]{acmart}
\usepackage{subcaption,enumitem} 
\usepackage{siunitx} 
\usepackage[normalem]{ulem}
\usepackage[skip=3pt,size=footnotesize]{caption}
\usepackage{algorithm}
\usepackage[noend]{algpseudocode}
\usepackage{amsmath}
\AtBeginDocument{%
  \providecommand\BibTeX{{%
    \normalfont B\kern-0.5em{\scshape i\kern-0.25em b}\kern-0.8em\TeX}}}

\newcommand{\name}{\mbox{POLAR}} 
\newcommand{\xref}[1]{\S\ref{#1}}

\newcommand{\textred}[1]{\textcolor{black}{#1}}
\ifx\noeditingmarks\undefined
   \newcommand{\pgwrapper}[2]{\textred{#1: #2}}
\else
   \newcommand{\pgwrapper}[2]{}
\fi

\newcommand{\textredd}[1]{\textcolor{black}{#1}}
\ifx\noeditingmarks\undefined
   \newcommand{\pgwrapperr}[2]{\textredd{#1: #2}}
\else
   \newcommand{\pgwrapperr}[2]{}
\fi

\newcommand{\cutt}[1]{}

\newcommand{\cut}[1]{}
\acmYear{2023}\copyrightyear{2023}
\setcopyright{rightsretained}
\acmConference[ACM MobiCom '23]{The 29th Annual International Conference on Mobile Computing and Networking}{October 2--6, 2023}{Madrid, Spain}
\acmBooktitle{The 29th Annual International Conference on Mobile Computing and Networking (ACM MobiCom '23), October 2--6, 2023, Madrid, Spain}
\acmPrice{}
\acmDOI{10.1145/3570361.3592504}
\acmISBN{978-1-4503-9990-6/23/10}
\setlist[itemize]{leftmargin=0pt,itemsep=0pt,parsep=0pt, wide= 0.01\parindent} %o/w *
\setlist[enumerate]{leftmargin=0pt, wide=0.01\parindent}

\begin{document}

\title{A Handheld Fine-Grained RFID Localization System with Complex-Controlled Polarization}


% \author{Anonymous Authors}
% \affiliation{Anonymous Affiliation}
% Use here ------------------------------
\author[Laura Dodds, Isaac Perper, Aline Eid, Fadel Adib]{Laura Dodds$^1$, Isaac Perper$^{1,3}$, Aline Eid$^{1,2}$, Fadel Adib$^{1,3}$}

\def \authors{Laura Dodds, Isaac Perper, Aline Eid, Fadel Adib}

\email{ldodds@mit.edu, iperper@mit.edu, alineeid@umich.edu, fadel@mit.edu}

\affiliation{\institution{$^1$ Massachusetts Institute of Technology, $^2$ University of Michigan, $^3$ Cartesian Systems}\country{}}

\renewcommand{\shorttitle}{\name}


\begin{abstract}
There is much interest in fine-grained RFID localization systems. Existing systems for accurate localization typically require infrastructure, either in the form of extensive reference tags or many antennas (e.g., antenna arrays) to localize RFID tags within their radio range. Yet, there remains a need for fine-grained RFID localization solutions that are in a compact, portable, mobile form, that can be held by users as they walk around areas to map them, such as in retail stores, warehouses, or manufacturing plants. 

We present the design, implementation, and evaluation of \name, a portable handheld system for fine-grained RFID localization. Our design introduces two key innovations that enable robust, accurate, and real-time localization of RFID tags. The first is \textit{complex-controlled polarization (CCP)}, a mechanism for localizing RFIDs at all orientations through software-controlled polarization of two linearly polarized antennas. The second is \textit{joint tag discovery and localization (JTDL)}, a method for simultaneously localizing and reading tags with zero-overhead regardless of tag orientation. Building on these two techniques, we develop an end-to-end handheld system that addresses a number of practical challenges in self-interference, efficient inventorying, and self-localization. Our evaluation demonstrates that \name\ achieves a median accuracy of a few centimeters in each of the x/y/z dimensions in practical indoor environments.

\begin{figure}[t]
    \centering
    \includegraphics[width=0.46\textwidth]{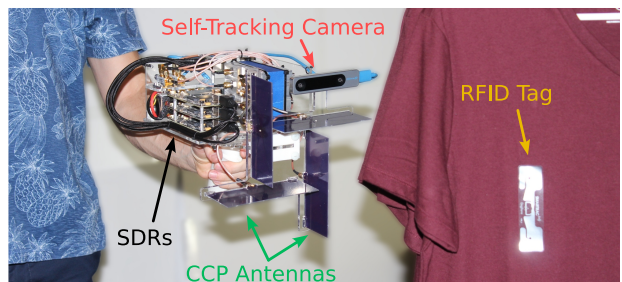}
    \caption{\footnotesize{\name\ Prototype.} \textnormal{Our handheld prototype consists of complex-controllable antennas, software radios with self-interference cancellation, a self-localization camera, and an onboard processor.}}
    \label{fig:cartesian}
    \vspace{-0.25in}
\end{figure}
\end{abstract}

% \vspace{-0.3in}




% \vspace{-0.3in}


\maketitle

%\everypar=\expandafter{\the\everypar\loosness=-1 }
%\everypar{\looseness=-1}

\vspace{-0.15in}
\section{Introduction} \label{sec:intro}
\vspace{-0.05in}

Fine-grained RFID localization has attracted much attention from the mobile and sensor computing community, due to its numerous applications spanning retail, manufacturing, warehousing, entertainment, and more~\cite{RFind,RFusion,PinIt,tagoram,MobiTag, LANDMARC}. In this paper, we set out to build a fine-grained handheld RFID localization system. We envision that a user of our system can walk around a typical indoor environment, such as a warehouse or a retail store, carrying a portable device that scans and accurately localizes all RFID-tagged items in its vicinity with fine-grained accuracy (10-20 centimeters). Such a capability would allow retailers, manufacturers, and warehouse operators to map all RFID-tagged items in their buildings and create digital twins of their environments to provide new analytics and make operational processes such as search, retrieval, and putaway more efficient.

\begin{figure*}[t]
    \centering
    \begin{minipage}[t]{0.44\textwidth}
        \begin{subfigure}[t]{0.49\linewidth}
            \centering
            \includegraphics[width=\linewidth,trim={0 3cm 0 5cm},clip]{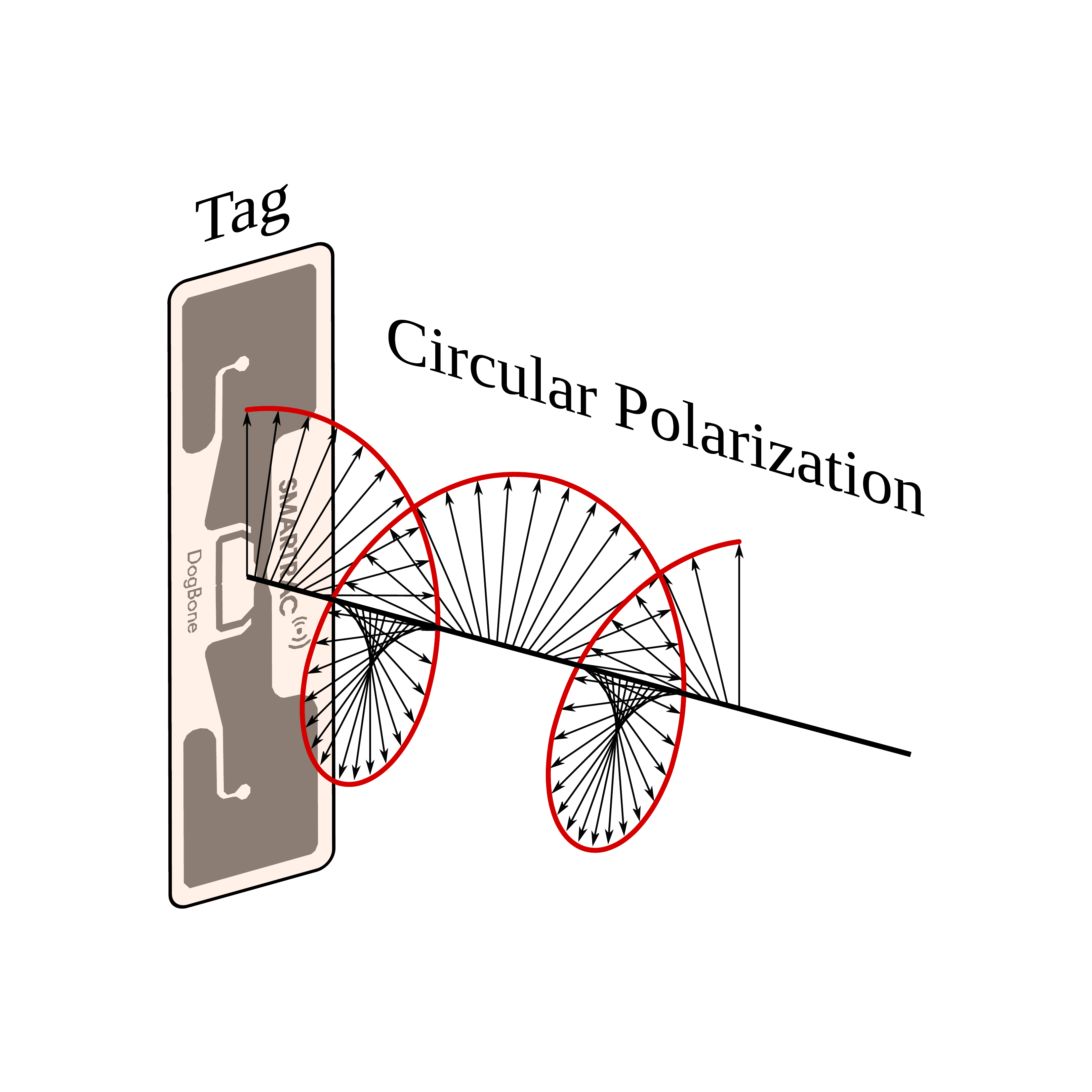}
            \vspace{-0.3in}
            \caption{\footnotesize{Phase of Vertical Tag.}}
            \label{fig:phase_horizontal}
        \end{subfigure}
        \begin{subfigure}[t]{0.49\linewidth}
            \centering
            \includegraphics[width=\linewidth,trim={0 3cm 0 6cm},clip]{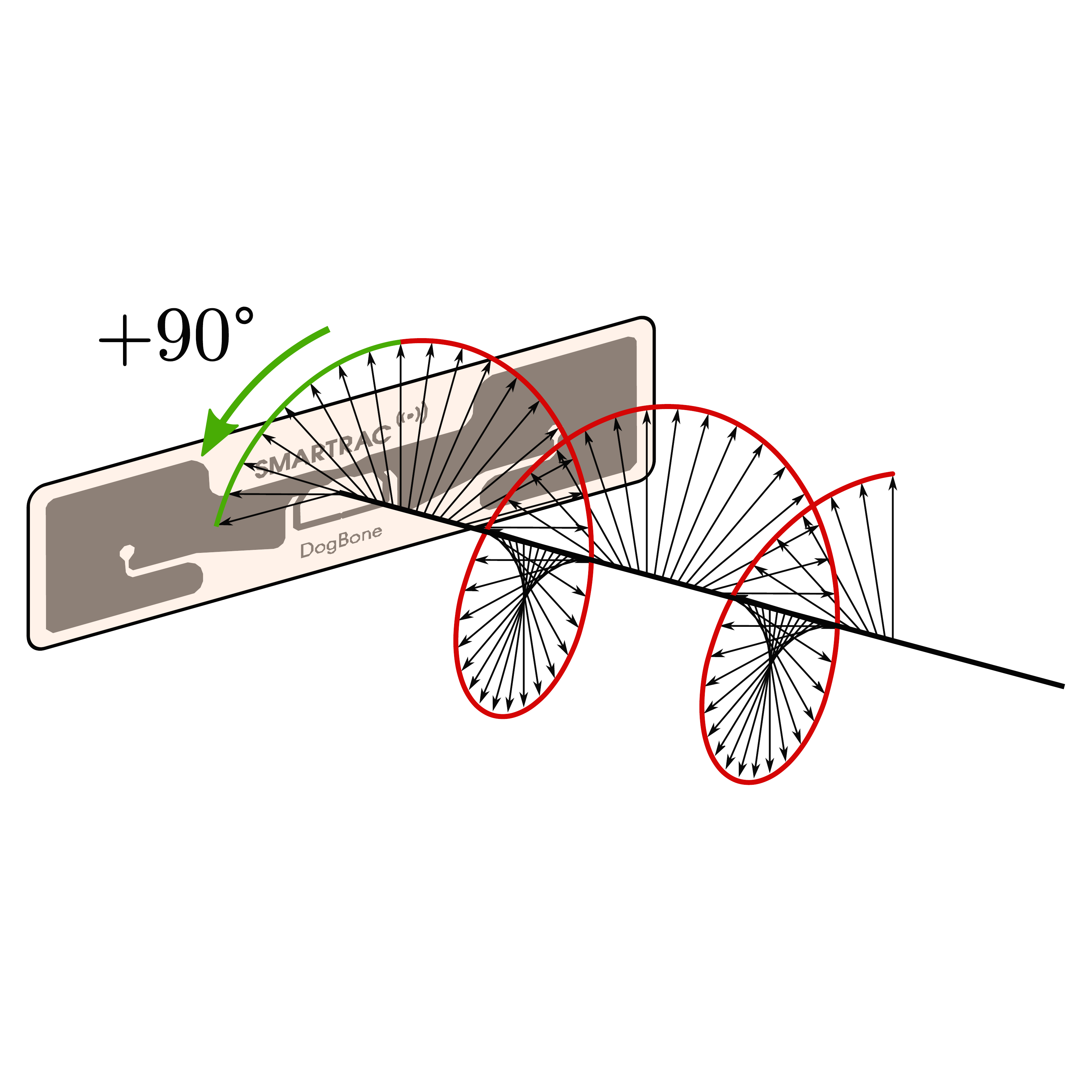}
            \vspace{-0.3in}
            \caption{\footnotesize{Phase of Horizontal Tag.}}
            \label{fig:phase_verticalintro}
        \end{subfigure}
        \vspace{-0.05in}
        \caption{\footnotesize{Impact of Tag Rotations. }\textnormal{(a) Tag reflects a CP signal when aligned with its polarization. (b) Rotated tag reflects at a different point, creating a phase offset. }}
        \label{fig:circ_phase}
    \end{minipage}
    \hspace{0.01\linewidth}
    \begin{minipage}[t]{.54\textwidth}
    
        \begin{subfigure}[t]{0.44\linewidth}
        
            \centering        \includegraphics[width=\linewidth,trim={0 0 0 0cm},clip]{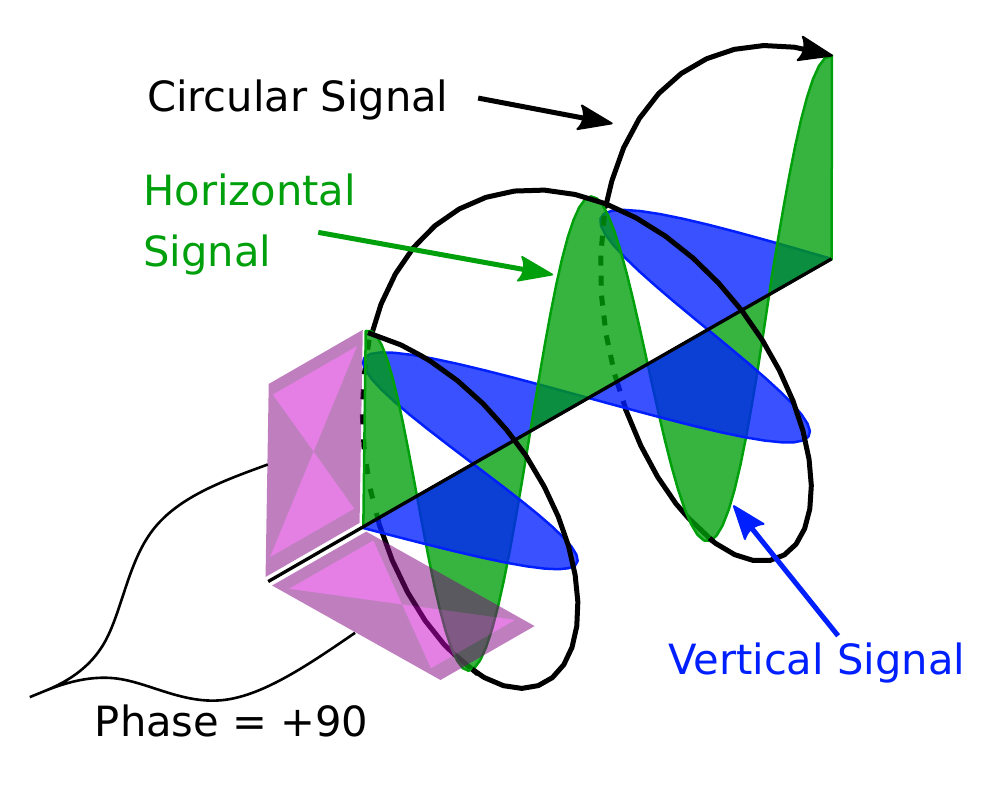}
            \vspace{-0.25in}
            \caption{\footnotesize{Constructing CP Signals.} } 
            \label{fig:circ_combine}

        \end{subfigure}
        \begin{subfigure}[t]{0.54\linewidth}
            \centering
            \includegraphics[width=\linewidth,trim={0 0 0 0cm},clip]{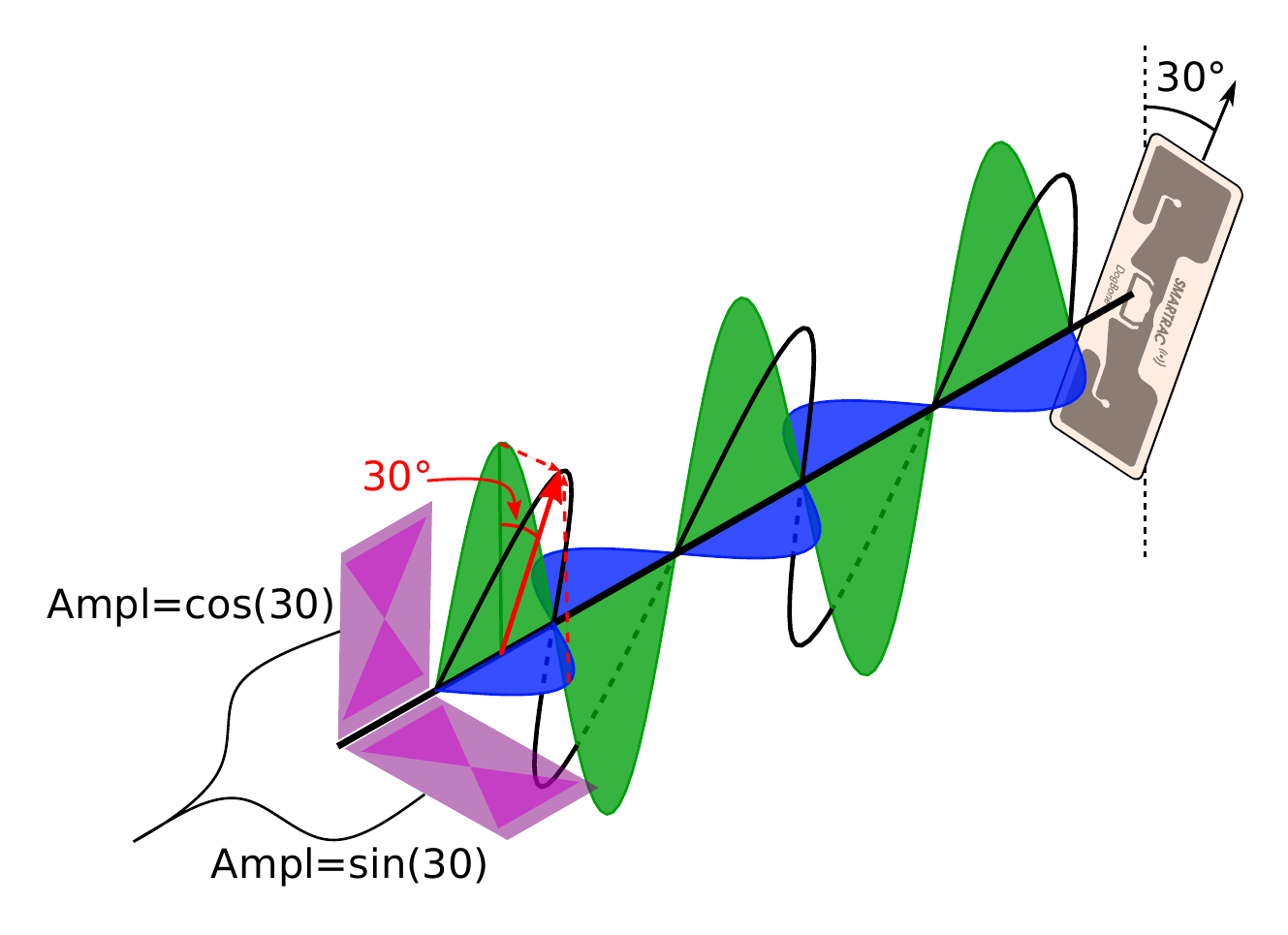}
            \vspace{-0.25in}
            \caption{\footnotesize{Constructing LP Signals.}}
            \label{fig:lin_pol_1}
        \end{subfigure}
        \vspace{-0.035in}
        \caption{\footnotesize{Complex-Controlled Polarization. }\textnormal{(a) 2 perpendicular signals with a 90\textdegree\ phase shift create a CP signal. (b) 2 perpendicular signals construct a 30\textdegree\ LP signal. }}
        
    \end{minipage}
    \vspace{-0.17in}
\end{figure*}

Unfortunately, state-of-the-art RFID systems cannot achieve the vision of fine-grained localization in a handheld form-factor. Most existing systems fall in two categories. The first consists of fine-grained RFID localization systems which leverage bulky antenna setups~\cite{PinIt,RFIDraw,RFind,ImpinjXarray}; these can achieve high accuracy but need to be deployed as an infrastructure with antennas spaced by meter-scale distances, making them unsuitable to be held by a user. The second category consists of commercial handheld RFID readers from companies like Zebra, Bluebird, and AsReader~\cite{ZebraHandheld, BluebirdHandheld, AsReaderHandheld}; these can be easily held by a user but cannot accurately localize RFIDs as they only detect the presence of tags, not their location. And while there have been proposals to leverage mobility in RFID localization~\cite{tagoram,MobiTag,RFusion,MobiTag,PinIt,TurboTrack}, most of these proposals require either moving the reader (or the tag) on predefined trajectories (e.g., by a robot), making them ill-suited for handheld human mobility.

We present \name\footnote{Polarization-based Orientation-independent Localization of Any RFID}, a handheld system for fine-grained RFID localization, shown in Fig.~\ref{fig:cartesian}. As a user carrying our device walks around an indoor environment, the device can self-localize, as it reads and localizes RFIDs within its radio range. This information allows it to create digital twins of indoor environments that are populated with the tagged items in 3D space. In contrast to prior approaches for fine-grained RFID localization which are constrained by the radio range of a deployed infrastructure, \name's handheld device can cover any area a user walks in, making this approach significantly more cost-effective and scalable than large-scale antenna deployments.

A major challenge we faced in developing \name\ was achieving fine-grained localization independent of an RFID's orientation. In practical deployments, RFIDs may be tagged at any orientation on their target items. Existing portable RFID readers typically rely on circularly polarized antennas, which allow them to power up and read tags irrespective of their orientation. However, the circular polarization adds an intractable phase to the RFID's channel if the tag orientation changes, even if the RFID remains in the same location; this makes accurate localization difficult.\footnote{This phase-based issue does not occur in antenna arrays (or SAR-based systems) because their localization is based on phase differences between antenna elements, allowing them to cancel out the orientation phase.} State-of-the-art localization relies on accurate phase measurements, meaning that phase distortions would make accurate localization very challenging. In principle, one could replace circularly polarized antennas with linearly polarized ones, but these cannot read tags that are in orthogonal (or near orthogonal) orientations, which could lead to missing up to half the tags in the environment and is the reason why portable readers rely on circularly polarized antennas in the first place.

To see why localizing RFIDs using circularly polarized antennas suffers from an unknown phase offset, consider the illustrative example shown in Fig.~\ref{fig:circ_phase}. In circularly polarized transmissions, the electric field (shown as the black arrows) rotates as the signal travels in space. Since RFIDs are linearly polarized,\footnote{Circularly polarized RFIDs exist, but are more expensive and larger than linearly polarized ones. Hence, they have significantly less adoption.} they power-up (and respond) when the received electric field is aligned with their orientation. Thus, as shown in Fig.~\ref{fig:phase_horizontal}, a vertical RFID tag would reflect the signal when the electric field is vertical. On the other hand, if the RFID is rotated, as shown in Fig.~\ref{fig:phase_verticalintro}, then a vertically-received electric field vector is not aligned anymore, and the signal needs to travel longer for the field to be aligned with the tag, indicated by the additional green arc. This adds a fictitious distance and creates a phase offset that is dependent on the RFID’s orientation.\footnote{Only rotation in the circular polarization plane impacts phase (see~\xref{sec:polarization}).}
This is why today's handheld readers, which leverage circularly polarized antennas, are incapable of performing accurate phase-based localization.

So how can we read \textit{and} accurately localize RFID tags independent of their orientation? We introduce a technique called \textit{complex-controlled polarization} (CCP). Our approach relies on two orthogonal, linearly polarized antennas with independent phase and amplitude control. Let us first show how \name\ can read RFIDs across orientations. Rather than using a single circularly polarized antenna, \name\ generates a circularly polarized signal by feeding the same signal to both linearly polarized antennas, with a 90\textdegree\ phase shift between them, as shown in Fig.~\ref{fig:circ_combine}. This allows \name\ to generate a circularly polarized signal using two linearly polarized antennas, and power up the RFID at any angle.

Next, let us see how this approach can localize tags across orientations. Recall that using circularly polarized antennas would introduce an unknown phase offset, which limits their localization accuracy. If the tag is vertical or horizontal, then one can simply transmit along the vertically-polarized or horizontally-polarized reader antenna, respectively. However, if the tag is at 30\textdegree\ as shown in Fig.~\ref{fig:lin_pol_1}, then if we use the vertical antenna alone, it would receive only a fraction of the power. Ideally, we would like to generate a linearly polarized signal whose orientation is aligned with the tag. To do so, we perform independent amplitude control across the two antennas. Specifically, rather than transmitting from only one of the antennas, we transmit the same signal but with $\cos(\pi\frac{30}{180})$ \textit{amplitude} along vertical antenna and $\sin(\pi\frac{30}{180})$ along horizontal antenna (with no phase offset), as shown in Fig.~\ref{fig:lin_pol_1}. This allows generating a linearly polarized signal at the corresponding angle (30\textdegree), matching that of the tag. This approach can be applied at any angle, allowing \name\ to achieve the highest signal-to-noise ratio (SNR) across orientations, and receive a response without a  phase offset, enabling accurate localization.

Building the above technique into an efficient and portable localization system faces a number of practical challenges. While CCP allows generating different polarizations for reading and localization, we would like to perform both processes simultaneously (rather than serially) to make the system efficient. The overall device also needs to be compact, ideally by using the same antennas for reading and localization. This is complicated by the need to realize simultaneously different polarizations and to deal with the self-interference across all simultaneous transmissions (and polarizations).

To address these challenges, \name\ introduces a mechanism called \textit{joint tag discovery and localization} (JTDL). Building on past work in dual-frequency excitation, the system decouples powering the RFIDs from localizing them~\cite{RFind,RFChord}. Specifically, our JTDL approach  generates signals to power the tags (in the UHF ISM band) in a circularly polarized fashion while transmitting localization frequencies outside the ISM in a linearly polarized manner. Our design takes this idea a step further by transmitting different frequencies at different times from the horizontal and vertical antennas and combining them in post-processing to emulate different polarizations and synthesize any orientation. We detail this technique in~\xref{sec:jtdl}, how \name\ performs self-interference cancellation in~\xref{sec:isolation}, and enable accurate 3D localization in~\xref{sec:3d}.

We built a prototype of \name\ (Fig.~\ref{fig:cartesian}). Our prototype uses an Intel Realsense T265~\cite{camera} to self-localize, and implements the algorithms for RFID discovery, identification, and localization on bladeRF software radios~\cite{bladeRF}. We also designed and fabricated compact wideband antennas with desirable gain and bandwidth properties for wideband RFID localization. 
Our empirical evaluation demonstrates:
\vspace{-0.05in}
\begin{itemize}
\item \name's CCP for circularly polarized transmissions lets it power up and read tags across all orientations. In comparison, for tags placed at the same distance, a linearly polarized antenna cannot power (or read) up to 30\% of tag angles.
\item \name's CCP-based approach for linearly polarized transmissions is able to read the tag's phase without a phase shift, achieving a 90\textsuperscript{th} percentile phase error less than 0.2 radians in the presence of tag rotations up to 90\textdegree. In contrast, if we use a circularly polarized antenna, the tag's response has an unknown phase shift, ranging anywhere from 0.03 to 3.1 radians (roughly $\pi$ radians, as expected for a 90\textdegree\ rotation). 
\item  Our end-to-end system can localize RFID tags with a median accuracy of 9cm and $90^{th}$ percentile of 17cm, indepen- dent of the tag's orientation, as a user carrying the device performs normal RFID inventorying, thus achieving accurate and efficient localization in a handheld form factor.
\end{itemize}
\vspace{-0.03in}
\noindent 
\textbf{Contributions:} The paper presents \name, a portable, handheld RFID localization system that is accurate, robust, and efficient. It introduces multiple innovations. First is complex-controlled polarization (CCP), an approach to localize RFID tags under random orientations using a handheld device with two linearly polarized antennas. Second is a mechanism to jointly perform discovery and time-of-flight estimation with zero overhead, independent of the tag orientation. The paper also contributes a proof-of-concept prototype implementation and experimental evaluation demonstrating the method’s accuracy and efficiency.

\vspace{-0.1in}
\section{Localizing Across All Angles} \label{sec:polarization}
\vspace{-0.03in}

In this section, we first describe the challenges associated with reading and locating tags at different orientations. Next, we show how \name's design overcomes these challenges.

\vspace{-0.1in}
\subsection{The Tag Orientation Problem} \label{sec:problem}
\vspace{-0.03in}

\begin{figure*}[t]
    \centering
    \begin{minipage}[b]{0.27\textwidth}
        \centering
        \includegraphics[width=0.75\linewidth]{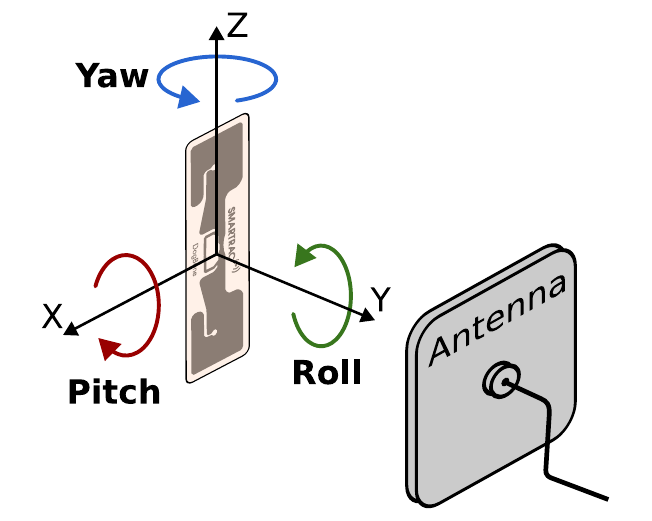}
        %\vspace{-0.1in}
        \caption{\footnotesize{Tag Rotations.} \textnormal{Axes of rotations: pitch (red), roll (green), yaw (blue).}}
        \label{fig:tag_rotations}
        \vspace{-0.1in}
    \end{minipage}
    \hspace{0.05in}
    \begin{minipage}[b]{.7\textwidth}
        \centering
        \begin{subfigure}[b]{0.32\linewidth}
            \centering
            \includegraphics[width=\linewidth]{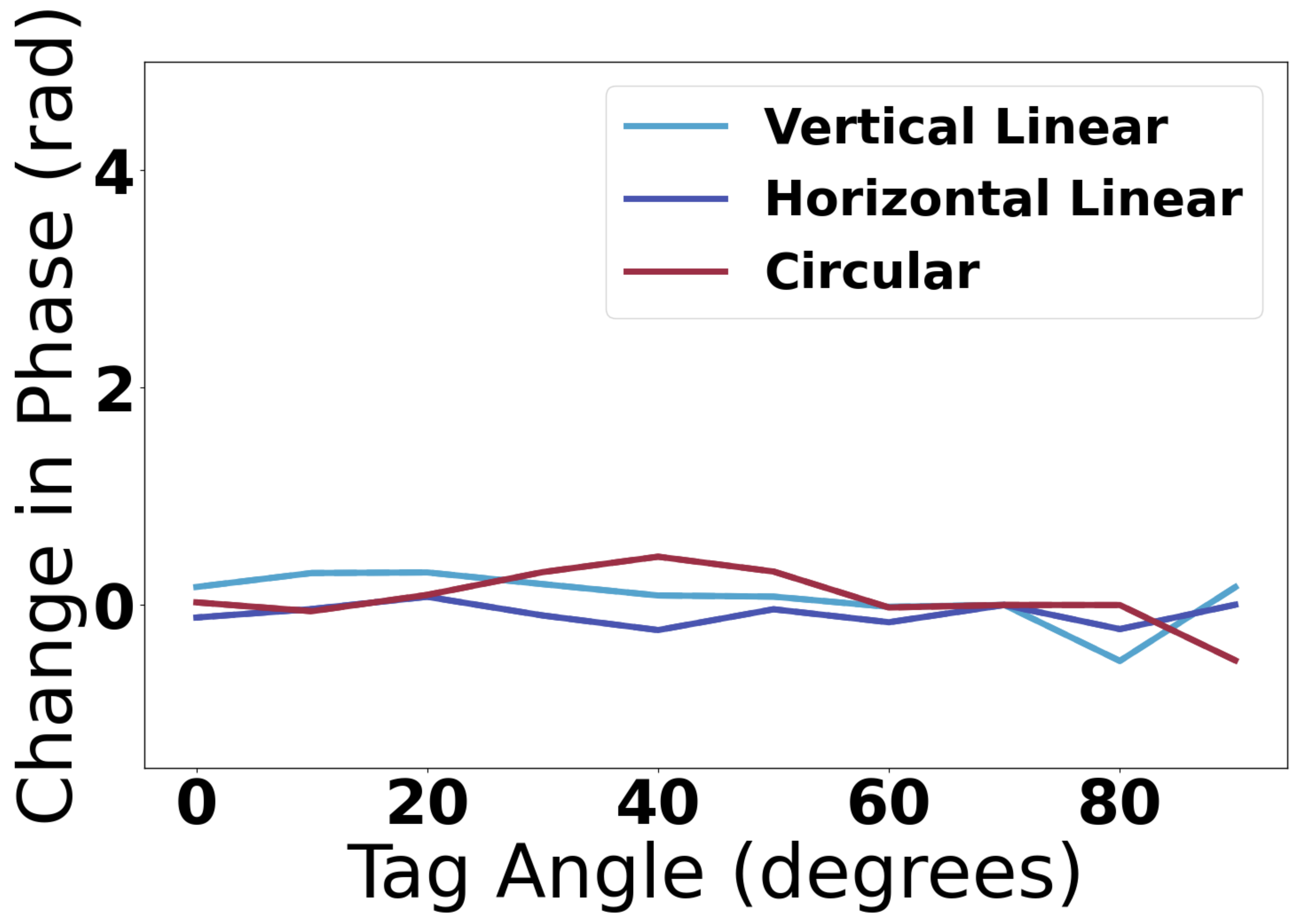}
            \vspace{-0.2in}
            \caption{\footnotesize{Pitch.}}
            \label{fig:pitch_x}
        \end{subfigure}
        \begin{subfigure}[b]{0.32\linewidth}
            \centering
            \includegraphics[width=\linewidth]{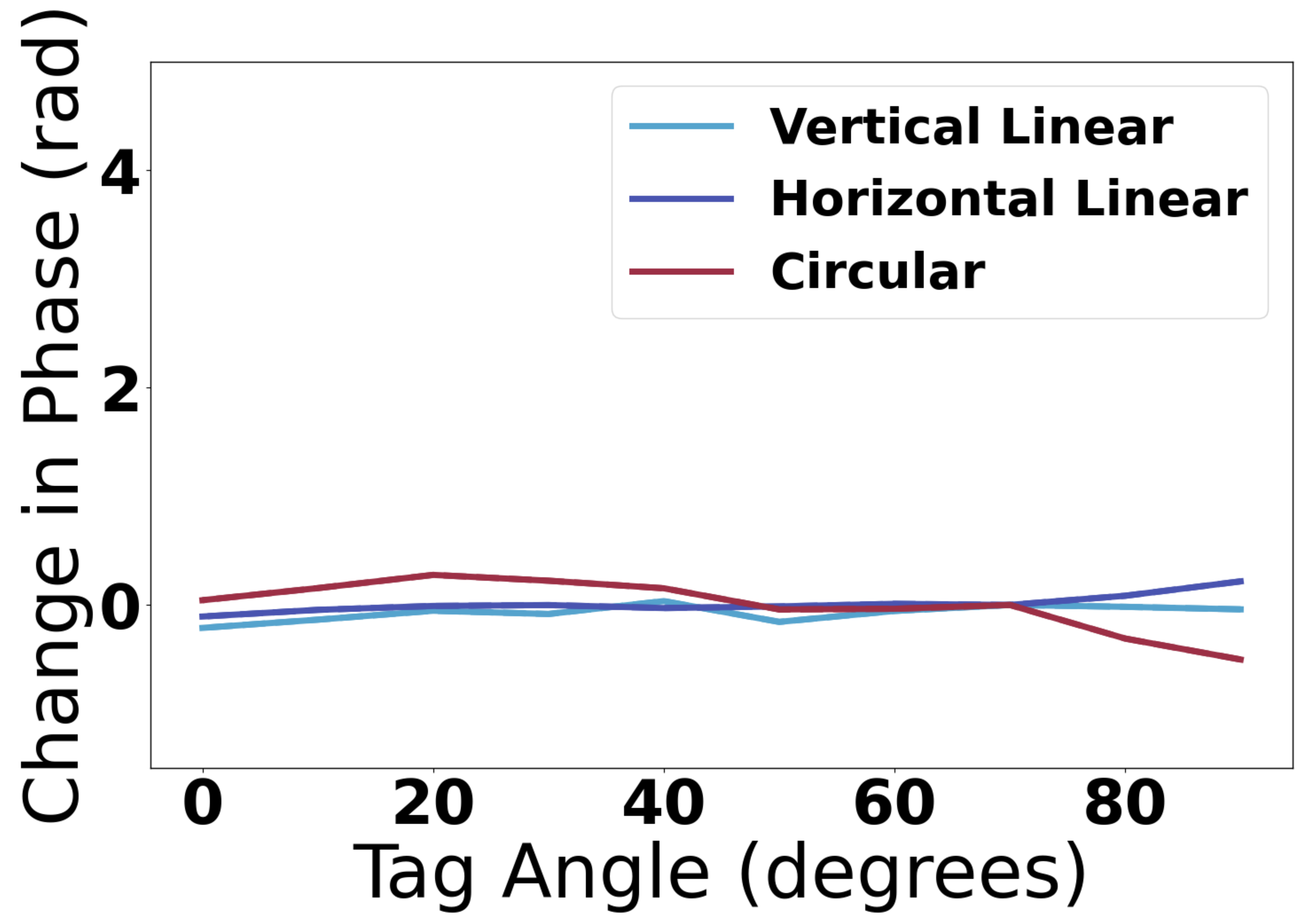}
            \vspace{-0.2in}
            \caption{\footnotesize{Yaw.}}
            \label{fig:yaw_z}
        \end{subfigure}
        \begin{subfigure}[b]{0.32\linewidth}
            \centering
            \includegraphics[width=\linewidth]{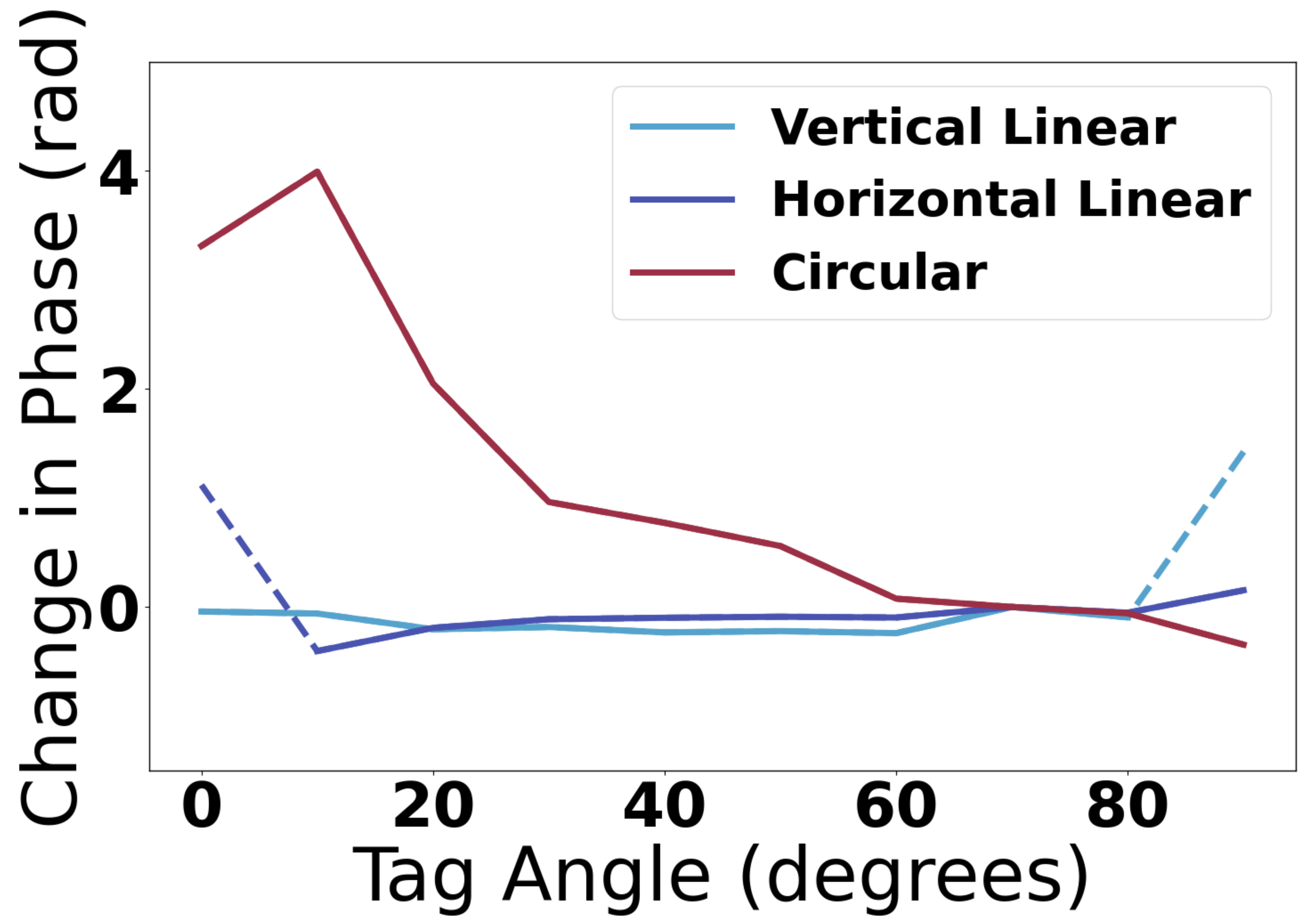}
            \vspace{-0.2in}
            \caption{\footnotesize{Roll.}}
            \label{fig:roll_y}
        \end{subfigure}
        \vspace{-0.03in}
        \caption{\footnotesize{Impact of Tag Rotations. }\textnormal{Change in phase of tag's response vs orientation, for a horizontally polarized (purple), vertically polarized (blue), and CP (red) antenna for (a) pitch, (b) yaw, and (c) roll. }}
        \label{fig:phase}
        \vspace{-0.1in}
    \end{minipage}%
    \vspace{-0.05in}
\end{figure*}

The majority of RFID readers rely on circularly polarized (CP) antennas because they can power and read tags regardless of orientation. However, in the presence of arbitrary tag rotations, these antennas can fail entirely in localization. Recall from~\xref{sec:intro} that the issue arises from the fact that most accurate localization systems rely on the phase of the tag's response to measure the distance to the tag. However, the phase for a CP signal depends on the orientation of the tag. In fact, certain tag rotations can cause phase offsets up to $\pi$, resulting in localization errors up to several meters.\footnotemark

To understand the impact of tag orientations, consider the three different rotations: roll, pitch, and yaw in Fig.~\ref{fig:tag_rotations}. Changing the tag's roll adds a phase offset for CP signals, shown in Fig.~\ref{fig:circ_phase}. This is because the tag's strongest reflection occurs when the tag is aligned with the signal's electric field. Fig.~\ref{fig:phase_horizontal} shows a CP signal incident to a linearly polarized tag. The phase of the response is determined by the point where the electric field (black arrows) becomes aligned with the tag. When the tag is rotated to a new angle in Fig.~\ref{fig:phase_verticalintro}, the signal needs to travel a further distance to be aligned with the tag, as indicated by the additional green arc. This introduces a phase offset that is dependent on the rotation of the tag. In contrast, changing the tag's pitch or yaw does not induce a phase offset, since these rotations do not change the direction of the tag's polarization relative to the CP signal. 

\footnotetext{Some solutions (e.g., antenna arrays) work despite this phase offset, but require bulky setups that need wall mounting and cannot be handheld.}

We investigated this phenomenon in a real-world experiment by measuring the impact of tag rotation on the tag's phase. We placed the tag and our reader at a fixed distance and we rotated the tag in intervals of 10\textdegree\ in the 3 directions (roll, pitch, yaw). At each angle, we measured the average phase and SNR. We repeated the experiment with a vertically polarized (blue), a horizontally polarized (purple) and a circularly polarized (red) antenna. 

Fig.~\ref{fig:phase} plots the change in phase of the tag's response (relative to the tag’s response at an initial orientation) as a function of the tag's angle for each of the three rotation directions. We focus on the range of 0\textdegree\ to 90\textdegree, since the remaining angles follow the same pattern.  In these plots, a solid line indicates that the SNR at that angle was above a threshold(e.g., -0.25~dB) and the dotted lines indicate that the SNR was below that threshold. We make a few observations:

\vspace{-0.03in}
\begin{itemize}
    \item Fig.~\ref{fig:pitch_x} and Fig. \ref{fig:yaw_z} show that for pitch and yaw, respectively, the phase remains consistent (i.e., the change in phase remains near 0) for the circularly polarized signal. 
    \item Fig.~\ref{fig:roll_y} demonstrates the issue that arises for changes in the tag's roll. The phase changes drastically, from 3.2 radians to -0.4 radians, for the CP signal, despite the distance to the tag remaining fixed. This means that the total change in phase is 3.6 radians (more than $\pi$ radians).
    This large change would result in meters of error in phase-based localization.  
    \item In contrast, for linearly polarized signals, the phase remains the same across all three rotations (if the tag is powered up). This is expected since the relative rotation between the signal's electric field and the tag remains fixed over time. Therefore, the tag's phase does not change with rotation. 
\end{itemize}
\vspace{-0.03in}

This experiment demonstrates that replacing a CP   antenna with a linearly polarized(LP) antenna eliminates the unknown phase offset. However, an LP antenna cannot power tags at all angles. This is due to polarization mismatch, where the tag only harvests energy from the portion of a signal that is parallel with its polarization~\cite{RotateAntennas}. Thus, when the polarization is near-perpendicular to the tag, the tag cannot harvest enough energy to power.

To investigate the impact of  polarization mismatch on powering up a tag, we conducted an experiment by placing the tag and reader at a fixed distance and rotating the tag's roll angle by 10\textdegree. At each angle, we read the tag 30 times and measured the SNR. If the tag was not successfully read,\footnote{We declared a reading to be successfully if the checksum was correct} we capped the measurement SNR at a minimum of -30dB. 

Fig.~\ref{fig:ib_snr_ours} plots max tag SNR vs roll angle for vertically (purple) and horizontally (red) polarized antennas. Our remarks:

\vspace{-0.03in}
\begin{itemize}
    \item When the tag is orthogonal to the polarization, the antenna cannot read the tag (the SNR is -30dB). Specifically, the horizontal~(90\textdegree) antenna cannot read tags from 0\textdegree\ to 20\textdegree. Similarly, the vertical~(0\textdegree) cannot read from 70\textdegree\ to 90\textdegree. 
    \item When the tag angle is not orthogonal to the polarization, the antennas can read the tag. For example, the vertical antenna~(0\textdegree) can read tags with high SNR between 0\textdegree and 70\textdegree. 
\end{itemize}
\vspace{-0.03in}
This shows that polarization mismatch prevents an LP antenna from reading a tag when within 20\textdegree\ of orthogonality.

\vspace{-0.1in}
\subsection{Complex-Controlled Polarization}
\vspace{-0.03in}

So far, we have described the challenges of reading and localizing tags at different orientations. To overcome these challenges, \name\ introduces \emph{Complex-Controlled Polarization (CCP)}. Our technique leverages perpendicular, linearly polarized antennas with independent phase and amplitude control, as shown in Fig.~\ref{fig:cartesian}.\footnote{In our design, we used 2 transmit antennas and 2 receive antennas, but the design could be reduced from 4 to 2 antennas by using circulators.} The remainder of this section describes how CCP enables both reading and locating tags at all orientations. For simplicity, we discuss these two steps independently then describe how we combine them in~\xref{sec:jtdl}.

\vspace{-0.1in}
\subsubsection{Reading Tags with CCP} Recall from \xref{sec:problem} that CP signals can power tags at all orientations(but they add an unknown phase to localization). Instead of using a CP antenna, \name\ leverages its 2 LP antennas to generate a CP signal. 

\label{sec:circ}

\begin{figure*}[t]
    \centering
    \begin{minipage}[t]{.25\linewidth}
            \centering
            \includegraphics[width =\textwidth]{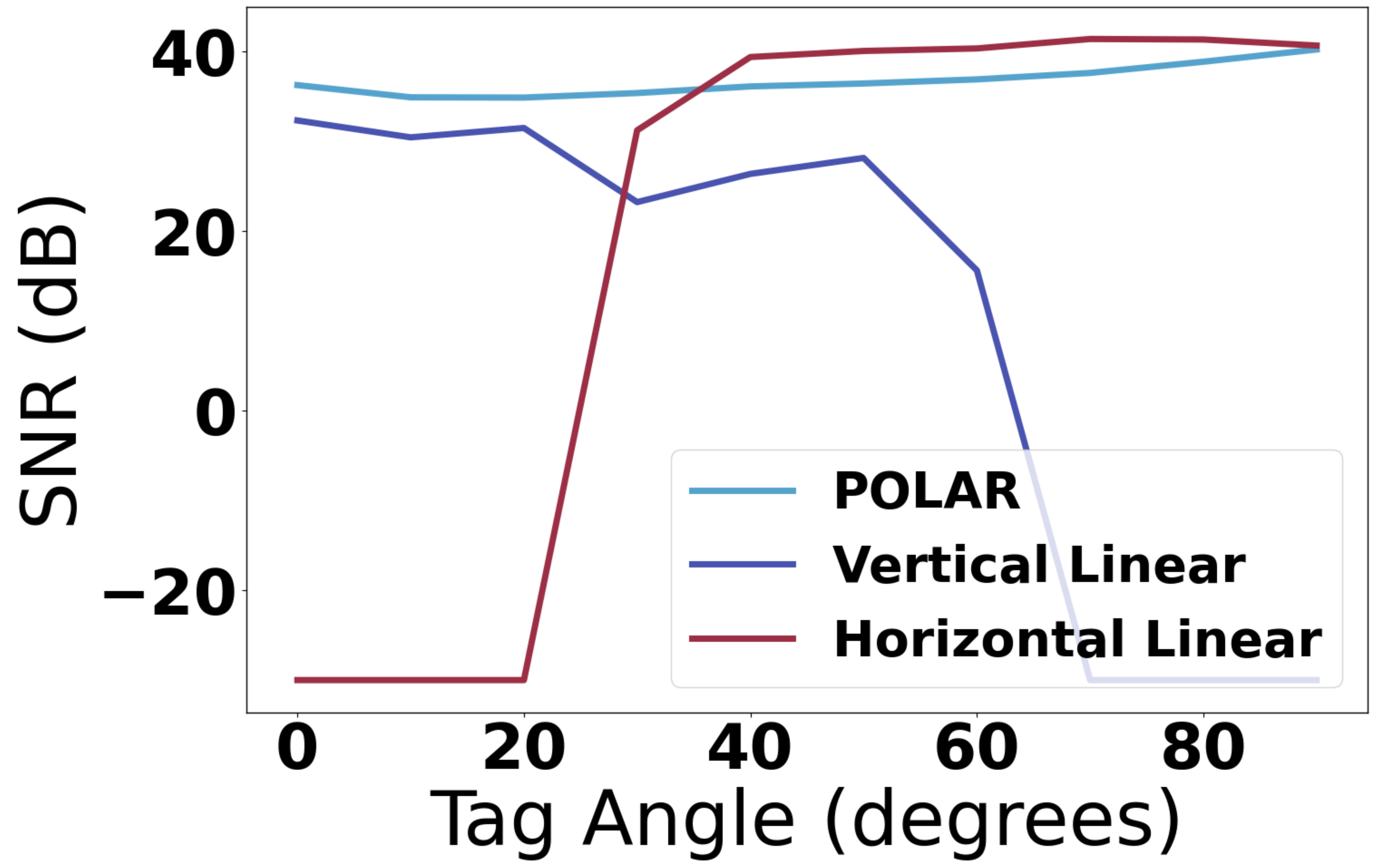}
            \vspace{-0.165in}
                \caption{\footnotesize{SNR vs Angle.} \textnormal{The plot shows SNR vs roll for \name\ (blue), horizontally polarized antenna (red), vertically polarized antenna (purple). }}
            \label{fig:ib_snr_ours}
    \end{minipage}
    \hspace{0.02in}
    \begin{minipage}[t]{0.25\textwidth}
            
        \centering
        \includegraphics[width=0.7\linewidth]{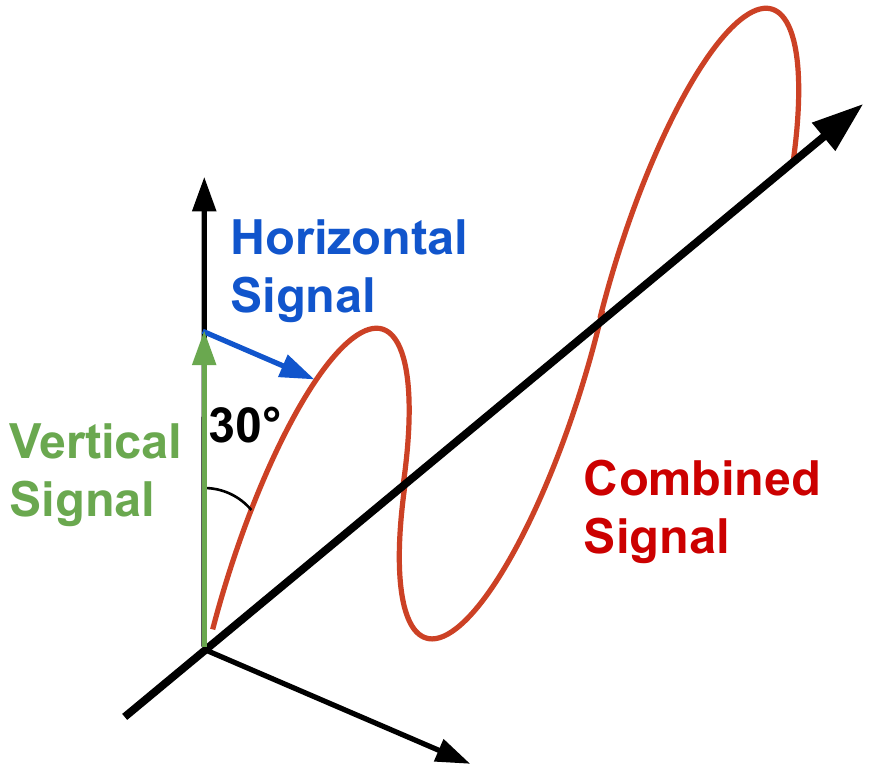}
        \caption{\footnotesize{Constructing Arbitrary Angles.} \textnormal{Two perpendicular signals (green,blue) with different amplitudes are sent. Combined signal is   sum(red).}}
        \label{fig:tilted_signal}
    \end{minipage}
    \hspace{0.02in}
    \begin{minipage}[t]{.225\textwidth}
            \centering
            \includegraphics[width=0.9\linewidth]{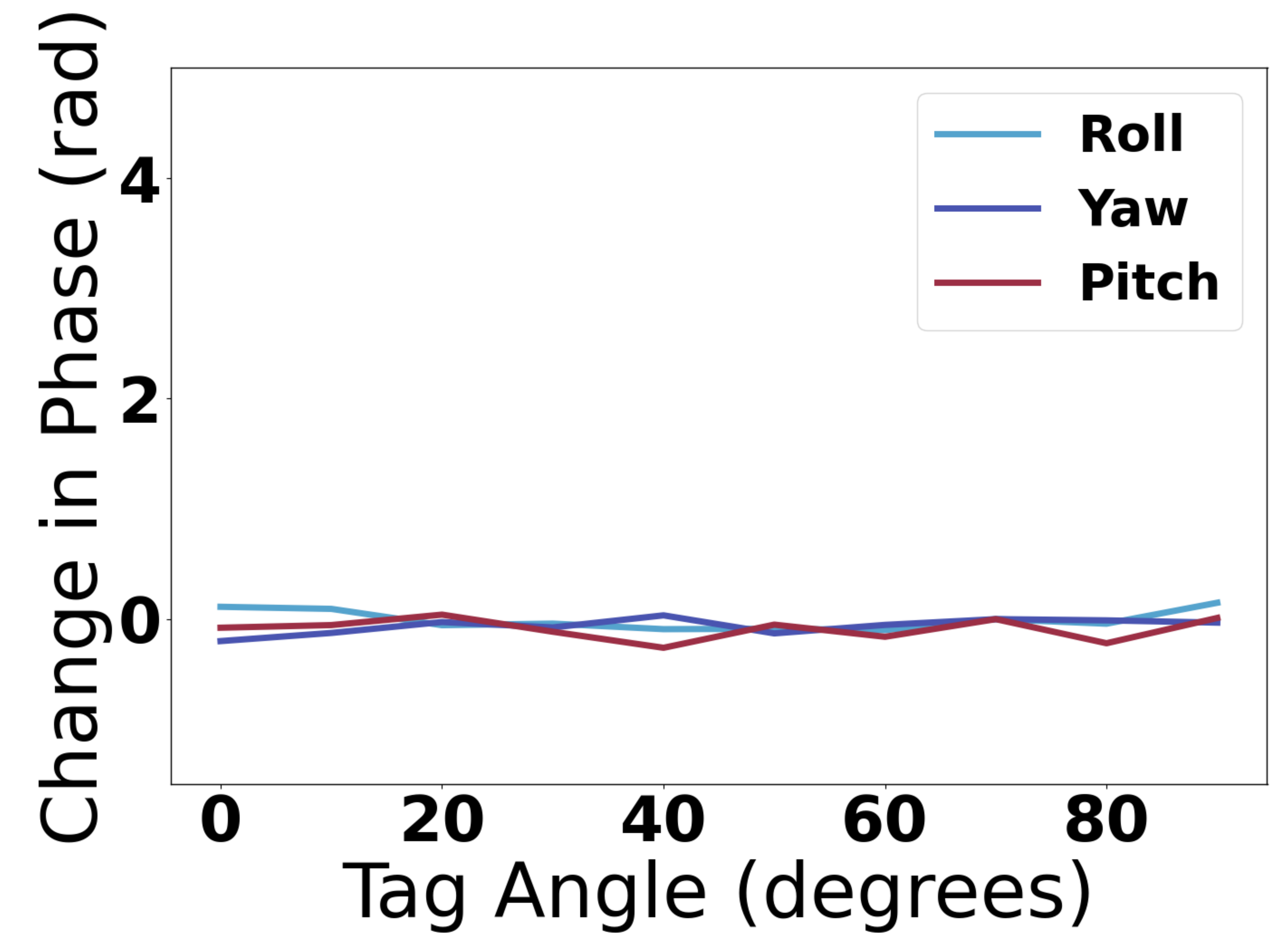}
            \caption{\footnotesize{\name's Phase.} \textnormal{Change in \name's phase vs angle when a tag rotates in roll(blue), yaw(purple), pitch(red).}}
            \label{fig:phase_ours}
    \end{minipage}%
    \hspace{0.02\linewidth}
    \begin{minipage}[t]{.225\textwidth}
            \centering
            \includegraphics[width=\linewidth]{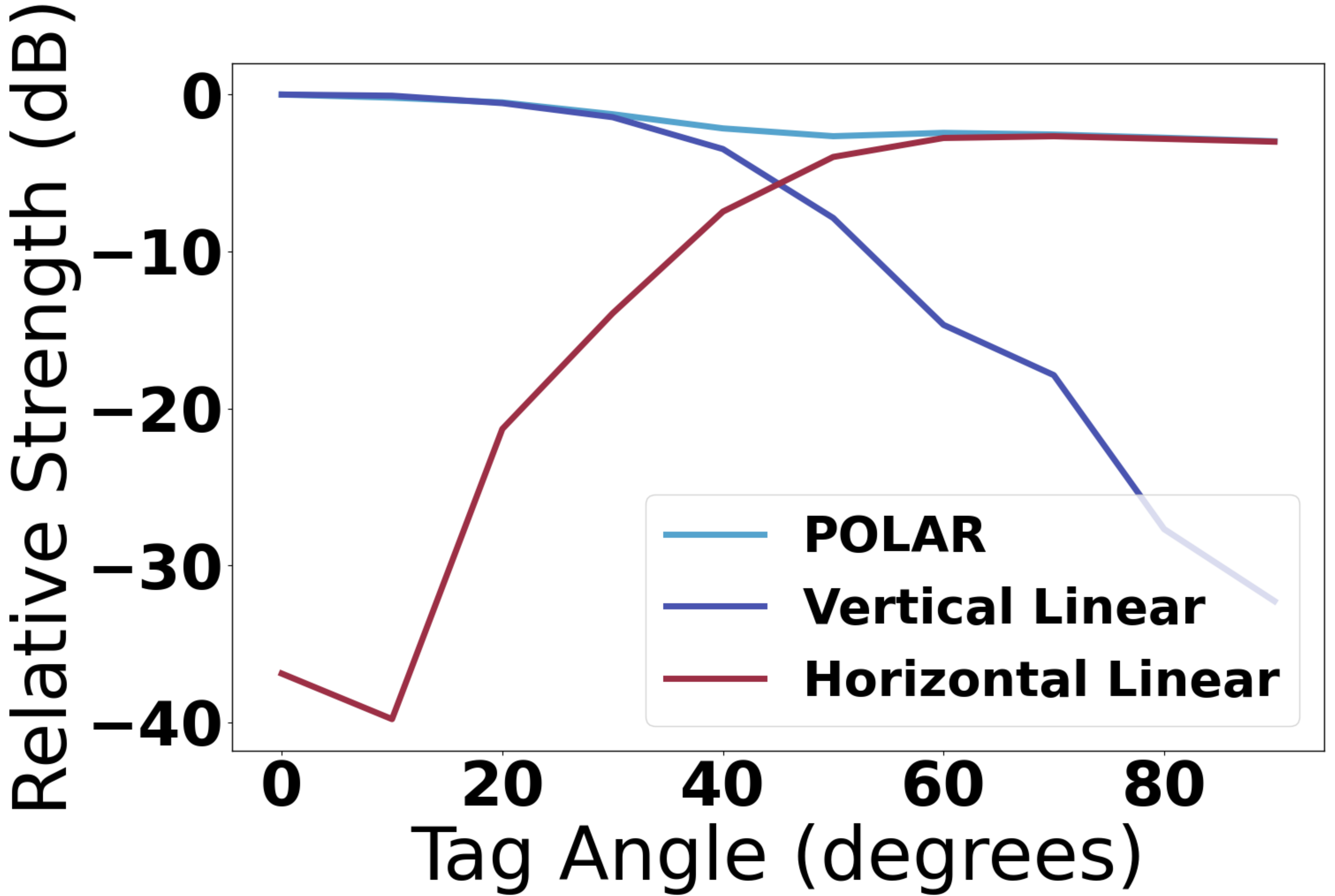}
            
            \caption{\footnotesize{RSS vs roll.} \textnormal{RSS vs tag roll angle using \name\ (blue), horizontally polarized(red), vertically polarized(purple) antenna. }}
            \label{fig:signal_strength}
    \end{minipage}
    \vspace{-0.15in}
\end{figure*}

Fig.~\ref{fig:circ_combine} shows how \name\ is able to construct the CP signal. \name\ sends a horizontal (blue) and a vertical (green) signal simultaneously, with a 90\textdegree\ phase shift between the two signals. At each point in time, the resulting signal is the vector addition of these two signals. As the signals travel, their phase offset causes this vector to rotate, creating a CP signal (traced out in black). This signal can power tags across orientations in the same way as a separate CP antenna. Formally, we send the following: 
{\small
\begin{equation}
    TX_{horiz,1} = x\ \ \ \ \& \ \ \ \ TX_{vert,1} = x e^{j \pi/2}
\end{equation}}
\vspace{-0.03in}

\noindent 
where $TX_{horiz,1}$ and $TX_{vert,1}$ are the transmitted signals on the two perpendicular antennas (horizontal and vertical), and x is the time-domain (modulated) RF signal. It is worth noting here that this approach is inherently different from typical MIMO (multi-input multi-output) systems in both theory and design~\cite{tse2005fundamentals}. This is because, in MIMO systems, antennas typically have the same polarization, and adding a phase offset does would not induce a time-varying polarization change of the transmitted signals.

Beyond powering the tag, we also need to decode the tag's response.
Similar to transmission, we synthesize a CP signal by combining the received signals with a 90\textdegree\ phase shift. This ensures that we can read the tag at all orientations.  

To demonstrate the effectiveness of this technique, we repeated the SNR experiment from~\xref{sec:problem} to measure the SNR of the tag’s response as a function of the tag’s roll. Fig.~\ref{fig:ib_snr_ours} shows the results for \name(blue). Unlike either of the LP signals, \name\ is able to achieve a high SNR(over 34dB) across all angles, which shows this design successfully creates CP signals that can power and read tags at all angles. 

\vspace{-0.065in}
\subsubsection{Localizing Tags with CCP} 
Next, \name\ employs an LP signal to localize the RFID. For simplicity, let us assume that the RFID has been powered up, and that our goal is to use  LP antennas to measure the phase of its response.%\looseness=-1

Recall that simply using LP signals results in polarization mismatch, causing a large drop in the SNR when the tag is near perpendicular to the signal. We observed this in Fig.~\ref{fig:roll_y}, where dashes lines denoted a very low SNR, when each of the horizontally and vertically polarized antennas were near orthogonal to the tag. Such SNR drop would impact channel estimates and lead to an inaccurate location estimate.

To overcome this, \name\ can construct an LP signal that aligns with the tag's orientation, minimizing losses from polarization mismatch.\footnote{Another option is to use many antennas with different polarizations, but that would lead to a bulky and expensive setup.} Fig.~\ref{fig:lin_pol_1} shows how \name\ constructs different LP signals. It simultaneously sends a vertical (green) and a horizontal (blue) signal. Unlike the CP signal described above, these two signals are sent with no phase shift. The resulting signal is their vector addition, shown in red, creating a linearly polarized signal at an angle. 

To change the orientation of this transmission, we change the relative amplitudes of the signals. Fig.~\ref{fig:tilted_signal} shows an example where the horizontal(blue) and vertical(green) signals are sent with different amplitudes. The combined signal(red) is at 30\textdegree. Therefore, through its independent amplitude control, \name\ can construct an LP signal at any angle. 

To generate a signal at a given angle $\theta$, \name\ needs to compute the necessary amplitudes. To do so, we note that the resulting signal is the hypotenuse of the right triangle formed by the two perpendicular signals. Therefore, to construct a signal at angle $\theta$, we can send:
{\small
\begin{equation}
TX_{horiz,2} = sin(\theta) x \ \ \ \ \& \ \ \ \ TX_{vert,2} = cos(\theta) x
\end{equation}

}

\noindent
where $TX_{horiz,2}$ and $TX_{vert,2}$ are the signals sent on the hor- izontal and vertical antennas, and $x$ is the modulated signal. 

Once the tag backscatters this signal, \name\ receives the response on its two perpendicular receive antennas. Each antenna will receive only the component of the tag’s response that is parallel to its polarization, again resulting in polarization mismatch. In theory, one could simply use the signal with the strongest response for localization. However, doing so would lose information from the other antenna, limiting the SNR. For example, when the tag is at 45\textdegree, each receive antenna will receive the same amount of power from the tag, so dropping the received signal from one antenna would result in losing half of the received power. 

Instead, \name\ combines the two responses to construct an LP receive signal that matches the tag’s orientation. This combination optimizes the power of the received signal, maximizing the SNR and therefore allowing accurate localization at further ranges.\footnotemark To combine these signals, we project the received signals onto an angle $\theta$ in a similar manner to the transmitted signal. Formally:
\begin{equation}
RX_{comb} = sin(\theta) RX_{horiz} + cos(\theta) RX_{vert}
\label{eq:rx_comb}
\end{equation}

\noindent
where $RX_{comb}$ is the combined signal, $RX_{horiz}$ and $RX_{vert}$ are the received signals on the horizontal and vertical antennas.

One might wonder whether adding a phase offset - rather than an amplitude offset - between the transmitted signals on the vertical and horizontal antennas may be a more appropriate approach for generating LP signals at different orientations. In practice, adding a phase offset would not lead to LP signals but rather CP ones. This is because a phase offset is equivalent to adding a delay between the transmitted signals, causing them to rotate with respect to each other over time. Indeed, our approach for generating a CP signal in~\xref{sec:circ} relied on a $90^\circ$ phase offset. Similarly, other phase offsets would result in elliptical polarizations with different major and minor axes (rather than linear polarizations).

\noindent
\textbf{Validating CCP.}
To evaluate this design, we assessed our ability to read both a consistent phase and a strong tag response, regardless of orientation. First, we investigated whether this design reads phase independent of orientation. We repeated the experiment from~\xref{sec:problem}, where the reader and tag were placed at a fixed distance and the tag was rotated in all 3 directions. We measured the change in phase (relative to the response at an initial orientation) for each rotation using \name's design. Fig.~\ref{fig:phase_ours} plots the change in phase as a function of the tag's roll (blue), yaw (purple), pitch(red). The range of the phase across all tag angles is below 0.2, 0.3, and 0.2 radians for roll, pitch, and yaw, respectively. These variations are minimal, and significantly smaller than those observed with the CP (specifically in Fig.~\ref{fig:roll_y}). The consistency of the phase across tag rotations demonstrates that our design reads orientation-independent tag phases, which is critical for accurate localization. 

\footnotetext{Proof omitted due to space constraints.}

Next, we investigated \name's ability to receive equivalent power across all tag rotations by using its CCP design to generate LP signals at any angle. We placed the tag and our reader at a fixed distance and rotated the tag's roll in intervals of 10\textdegree. To ensure the tag was powered for every trial, we placed separate antennas  for powering and reading the tag. We measured the channel strength at each angle. We compare this to the signal strength when using a single LP antenna, both horizontal and vertical.  

Fig.~\ref{fig:signal_strength} plots the normalized signal strength (relative to the max) vs tag roll for \name\ (blue), a horizontal antenna (red), and a vertical antenna (purple). For both horizontal (90\textdegree) and vertical (0\textdegree) antennas, the impact of polarization mismatch can be seen by a significant decline of over 25dB in signal strength as the tag moves closer to perpendicular. For \name, the signal strength remains consistent across all tag angles, varying by less than 3dB. This shows CCP's effectiveness in overcoming polarization mismatch.

\vspace{-0.1in}
 \section{Joint Discovery \& Localization} \label{sec:jtdl}
\vspace{-0.03in}
In the previous section, we described how \name\ leverages CCP to construct both circularly polarized signals for powering tags and linearly polarized signals for localizing tags. In this section, we describe how \name\ combines these techniques to simultaneously read and localize tags. 

A key goal is to localize tags \textit{during} a standard RFID inventorying process and without additional overhead for localization. To realize this goal, \name\ starts by building on a technique known as dual-frequency excitation~\cite{RFind}. In this technique, two signals of different frequencies are sent to the RFID tag: one high-power signal in the UHF ISM band to read the tag and one low-power signal for sensing. While the high-power signal must remain within the tag's narrow bandwidth to successfully read the tag, the sensing signal can be sent at any frequency. Since the RFID tag is frequency agnostic, it will reflect both signals. Thus, by varying the sensing frequency across a wide bandwidth, this technique can be used to measure ultra-wideband (UWB) channel estimates for accurate localization. 

Building on this technique, \name\ simultaneously sends its circularly polarized signal (to power the tag) and a linearly polarized signal (to localize). Since these signals are at different frequencies, we can send both from the same CCP antennas without them impacting each other's polarizations. To efficiently localize tags during discovery, we introduce a technique called \emph{Joint Tag Discovery and Localization} (JTDL), where we leverage the CCP technique to simultaneously perform tag discovery and UWB sensing, collecting over 200MHz of bandwidth within each tag read. 

\begin{figure}[t]
    \centering
    \includegraphics[width =0.475\textwidth]{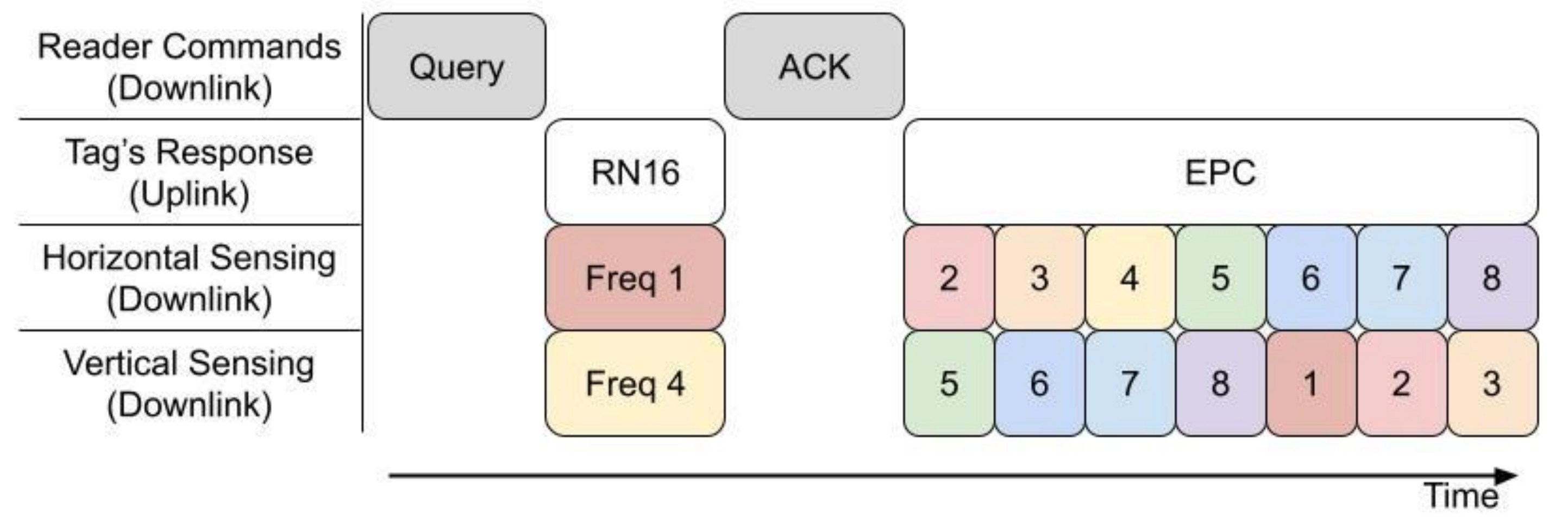}
    \caption{\footnotesize{Protocol Timing.} \textnormal{Timing of one JTDL round. The top 2 rows show the reader's downlink commands and tag's uplink response. The bottom 2 rows show the two LP sensing signals. Different frequencies (shown by different colors) are sent at different times to avoid interference.}}
    \label{fig:schedule}
    \vspace{-0.2in}
\end{figure}

One challenge with performing joint discovery and localization with CCP is that the angle of each tag is unknown a priori, making it infeasible for us to construct an LP transmit signal to match the tag’s angle as described in~\xref{sec:polarization}. 

To overcome this challenge, our idea is to send any given frequency on both antennas, but at different times. This allows measuring the horizontal and vertical components of the tag response, and combing them in post processing to achieve optimal SNR. 
The technique follows a 3-step process: 

\vspace{-0.03in}
\begin{enumerate} 

\item \textbf{Transmission.} \name\ transmits the horizontal and vertical LP signals at different times. However, separating the two transmit signals would require twice the transmission time, making it inefficient.
Instead, we transmits different frequency permutations on the 2 transmit antennas simultaneously. Fig.~\ref{fig:schedule} shows the schedule used to fit all frequencies within one round of the standard EPC Gen2 protocol~\cite{epcGen}. The first 2 rows show the reader’s downlink (CP) commands and the tag's uplink responses. During this process, \name\ measures the tag’s channel whenever the tag is backscattering (the RN16 and EPC). The bottom 2 rows show the transmitted frequencies for each of the LP antennas, with colors denoting different frequencies. Since the simultaneous frequencies are different, they do not interfere with each other and can be sent concurrently. With this, \name\ can measure >200MHz of bandwidth for each tag read, sufficient for accurate localization~\cite{RFind}.\footnote{If the same frequency is sent on both antennas, \cutt{they would result in}\textredd{the result would be} an LP signal at 45\textdegree, which causes large polarization mismatch for tags near -45\textdegree.} This is repeated until all tags within the device's radio range are read.

\item \textbf{Signal Combination. }Next, for each frequency, \name\ combines the channel measurements from each antenna in post-process to achieve optimal SNR. To do so, we simply perform maximal ratio combining~\cite{tse2005fundamentals}. 

\item \textbf{Distance Estimation. }Finally, with its UWB channel estimates, it can now invert the channel to estimate the time-of-flight to the tag and measure the 1D distance using the same methods applied in past work~\cite{RFind,TurboTrack,RFusion,RFChord}. \footnote{As demonstrated in past work\cite{RFind}, UWB enables time-of-flight estimation for multiple propagation paths, which allows identifying the direct path for accurate distance estimation in multipath.}
\end{enumerate}

\vspace{-0.03in}

\name\ repeats this process for every tag that it reads during discovery, which allows it to compute the 1D distance estimates for all tags in the environment accurately and efficiently. In~\xref{sec:3d}, we describe how \name\ leverages these 1D measurements to perform robust 3D localization.

\vspace{-0.1in}
\section{Addressing Self-Interference} \label{sec:isolation}
\vspace{-0.03in}

A well-known problem in designing compact RFID readers is self-interference~\cite{RFIDIntf}. Since these readers are full duplex (i.e., they transmit while receiving the  RFID response), the transmitted signals leak back to the receiver. This leakage is much stronger than the backscatter response and can overwhelm the receiver, preventing successful decoding. To deal with this leakage, RFID readers either separate the transmitter and receiver (e.g., by half a meter) or employ some self-interference cancellation scheme~\cite{FDrelay2,timo,WiVi}, whose choice depends on the reader implementation, bandwidth, antenna design, etc. Since we want to build a compact handheld reader, we cannot separate our antennas by a large distance, so we need to incorporate cancellation into our design.

Bringing self-interference cancellation to \name\ faces two key challenges. First, the system needs to not only cancel  self-interference from a typical RFID reader signal, but also from the UWB out-of-band signal, and it is more difficult to cancel wideband signals than typical narrowband RFID signals~\cite{fullduplex-new}. Second, \name\ has two different modes for transmission that happen simultaneously: the in-band is transmitted as a CP signal, and the out-of-band is transmitted as an LP. This adds further complexity, as we need to cancel two types of signals that are transmitted simultaneously.

To deal with self-interference, we sought simple and effective mechanisms, which we explain below.

\vspace{-0.1in}
\subsection{Dealing with CP Interference}\label{sec:circcancel}
\vspace{-0.03in}

\name\ aims to limit the self-interference from the CP signal used to power the tags. To do this, our approach is to leverage a method called \textit{cross-polarization} in the context of CP antennas. At a high level, cross-polarization means that if a transmitter and receiver have orthogonal polarizations, then the transmitted signal is significantly attenuated at the receiver. Indeed, we saw this in~\xref{sec:problem}, where a horizontally polarized antenna could not power up a vertical  tag. While this phenomenon is problematic in powering RFIDs, we harness it to cancel self-interference of our in-band signal.

To harness cross-polarization, we first note that our in-band transmitted UHF signal is CP. Specifically, we transmit with right-hand circular polarization (RHCP), i.e.,  the electric field is travelling clockwise, as in 
Fig.~\ref{fig:circ_combine}.
For two antennas with opposite polarization directions (i.e., facing each other), the cross-polarization of an RHCP transmission is typically left-hand circular polarization (LHCP), which rotates in the opposite direction, (i.e., counter-clockwise). However, in our design, the transmitter and receiver share the same polarization direction (i.e., they are both facing the same direction). Therefore, the cross-polarization of our RHCP transmitter is an RHCP receiver. 
This ensures that the transmission and reception remain orthogonal, minimizing the received signal.

Next, let us mathematically formalize this concept. Based on the earlier discussion, any polarization can be described using a 2D complex vector $\left[E_h, E_v\right]$, where the coordinates correspond to the horizontal and vertical (complex) numbers applied to the transmitted vector. An RHCP polarization can be realized as $E_{RHCP} = 1/\sqrt2[1,e^{-j\pi/2}]$, where $1/\sqrt2$ is the power normalization factor. Therefore, the received leakage signal in the ideal case is:
{\small
\begin{equation}\label{eq:polz}
    \left<E_{RHCP},E_{RHCP}\right> = \frac{1}{\sqrt{2}}\begin{bmatrix} 1,e^{-j\pi/2} \end{bmatrix} * 
    \frac{1}{\sqrt{2}}\begin{bmatrix} 1\\e^{-j\pi/2} \end{bmatrix} = 0
\end{equation}
}

To implement RHCP at the receiver using 2 linearly polarized antennas, we use a similar method as for generating the transmit signal. We add a 90\textdegree\ phase shift between the two received signals.

\begin{figure}[t]
    \centering
    \includegraphics[width=\linewidth, trim={0 0 0 1cm},clip]{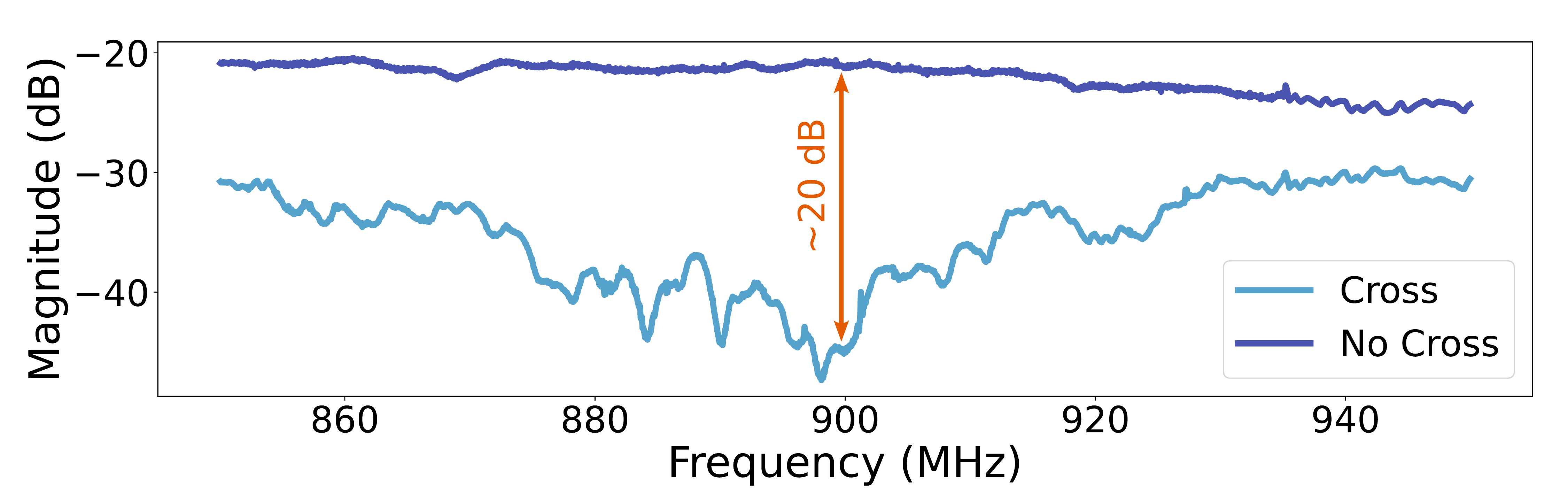}
    % \vspace{-0.2in}
    %  \vspace{-0.03in}
    \caption{\footnotesize{Impact of Cross-Polarization.} \textnormal{The plot shows the leakage for a CP signal with (blue) and without (purple) cross-polarization.}}
    \label{fig:isolation_cross}
     \vspace{-0.2in}
\end{figure}

To investigate the impact of our cross-polarization mechanism on cancelling the leakage, we \cutt{ran an experiment to }measure\textredd{d} the isolation (i.e., attenuation of the leakage) between the transmitter and receiver. To do so, we placed the setup in a large, open space with RF absorbers on the floor and covering all equipment (to mitigate the impact of reflections off the surrounding environment). Using a vector network analyzer, we measured the isolation between the transmitter and receiver from 850MHz to 950MHz (covering the international ISM bands). Fig.~\ref{fig:isolation_cross} plots the isolation with (blue) and without (purple) the cross polarization as a function of frequency. When receiving without cross polarization at 900~MHz, the design only achieves 21~dB of isolation.\footnote{This natural isolation is due to the antenna spacing.} In comparison, \name's design with cross-polarization achieves 45~dB of isolation, \cut{i.e., }an improvement of more than 20~dB (i.e., $100\times$).

One might wonder whether cross-polarization would not only mitigate self-interference, but also attenuate the received backscatter response. Thankfully, that is not the case; once the signal backscatters, it becomes linearly polarized. A CP antenna can receive an LP signal, regardless of its polarization. Indeed, that is why we used CP in the first place in our design of the in-band transmitter as described in~\xref{sec:polarization}.

We can formally see this by considering the polarization projections. If a tag's polarization is $E_{tag} = \left[\cos(\theta),\sin(\theta)\right]$, the amplitude of the signal received by the tag is\footnote{\cutt{Here, w}\textredd{W}e use the conjugate to account for the fact that the transmitter and receiver\cutt{ (tag)} have opposite polarization directions (\cutt{i.e., }they\cutt{ are}\textredd{'re} facing each other)~\cite{polarization_conjugate}.}:

{\small
\begin{equation}\label{eq:poltag}
\begin{split}
    \left<E_{RHCP},E_{tag}^*\right> = \frac{1}{\sqrt{2}} \begin{bmatrix} 1,e^{-j\pi/2} \end{bmatrix} * 
    \begin{bmatrix} \cos(\theta)\\ \sin(\theta) \end{bmatrix}
\\= 
   \frac{1}{\sqrt{2}}(\cos(\theta) + e^{-j\pi/2} \sin(\theta))
   \end{split}
\end{equation}
}

This signal is backscattered by the tag, which one can model as re-emitting the received signal along the same polarization\footnote{We maintain the same frame of reference\cutt{, and therefore again apply the conjugate to $E_{tag}$} \textredd{as Eq.~\ref{eq:poltag}.}}.  Hence, the polarization of the received signal can be expressed as:\footnote{Here, we ignore the amplitude and phase of backscatter as these are simply scalars from the perspective of polarization}

{\small
\begin{equation}\label{eq:poltagback}
\begin{split}
    \left<E_{tag}^*,E_{RHCP}\right> = \frac{1}{\sqrt{2}}\begin{bmatrix} \cos(\theta), \sin(\theta) \end{bmatrix} * 
    \begin{bmatrix} 1 \\ e^{-j\pi/2} \end{bmatrix} \\= 
    \frac{1}{\sqrt{2}}(\cos(\theta) + e^{-j\pi/2} \sin(\theta))
     \end{split}
\end{equation}
}
\vspace{-0.1in}

The above shows that even though the reader's transmit and receive are circularly cross-polarized, they are able to receive an LP backscatter signal. It is worth noting that this elegant concept has been observed and documented in the context of polarization light filters~\cite{wu2020turboboosting}.

\vspace{-0.1in}
\subsection{Dealing with LP Interference}  \label{sec:nulling}
\vspace{-0.03in}

In addition to the CP signals, \name\ needs to minimize the self interference from the LP signals used for localization. 
One might wonder if we can apply a self-interference cancellation mechanism similar to the cross-polarization that we used for the in-band CP signal. While a cross-polarized receive antenna (e.g., a horizontal receive antenna for a vertical transmit antenna) would attenuate the leakage, this would also significantly attenuate the tag's response. For example, consider a vertical tag. When transmitting on a vertical antenna, the tag's backscatter response will be strong. However, a horizontal antenna cannot receive its response due to the polarization mismatch (as demonstrated in~\xref{sec:problem}). Similarly, when transmitting on the horizontal antenna, the reflected signal will be weak due to polarization mismatch. In this case, \name\ would be unable to measure the tag response from either antenna pair. Thus, it must use parallel transmit and receive antennas, leading to high leakage.

To overcome this, \name\ employs over-the-wire nulling, building on past work in MIMO interference cancellation~\cite{tse2005fundamentals,WiVi}. At a high level, it estimates the leakage signal\footnote{\textred{Note that this can be a one-time calibration, since the self-interference is dominated by antenna leakage, which does not change over time.}} and injects another signal into the receiver, structuring it so that it destructively combines with the leakage at the receiver. (See~\cite{tse2005fundamentals} for more details.) 
\name\ repeats this process for each frequency for wideband nulling, and performs it independently for the vertical and horizontal antenna pairs. 

We investigated the impact of nulling by measuring the cancellation between \cutt{the }parallel antennas. We transmitted a fixed signal and compared the magnitude of leakage with and without nulling. 
We repeated this for all frequencies used in localization.

\begin{table}[t]
\centering
\footnotesize
\setlength{\tabcolsep}{5pt}
\begin{tabular}{ |c|c|c|c|c|c|c|c|c|c| }
 \hline
& \multicolumn{8}{c|}{\textbf{Frequency (MHz)}} & \\
\hline
& \textbf{763} & \textbf{790} & \textbf{833} & \textbf{871} & \textbf{915} & \textbf{938} & \textbf{973} & \textbf{1008} & \textbf{Avg}\\
\hline
 \textbf{Vert} & 16 & 22 & 31 & 23 & 29 & 21 & 21 & 22 & 23\\
 \textbf{Horiz}  & 21 & 21 & 31 & 22 & 26 & 19 & 23 & 18 & 23\\
 \hline
\end{tabular}
\caption{\footnotesize{\textbf{Nulling Cancellation (in dB).} \textnormal{The table shows the cancellation from wire nulling across frequencies for the Vert and Horiz antennas.}}}
\label{table:isolation_nulling}
\vspace{-0.15in}
\end{table}

\setlength{\floatsep}{16pt}
\setlength{\textfloatsep}{8pt}
\begin{algorithm}[t]
\footnotesize
\captionsetup{font=footnotesize,labelfont=footnotesize, labelfont=bf}

\caption{Measurement Selection Algorithm}
\begin{algorithmic}
\For {Tag in Tags}
    \State \textbf{INITIAL FILTERING}
    \For {$m_i$ in measurements}
        \If {$SNR_i < \tau$}
            \State Remove $m_i$
        \EndIf
    \EndFor %\\
    \State \textbf{MEASUREMENT SORTING}
    \State Compute bounding box of all measurements
    \State Split bounding box evenly into a grid of 3x3x3
    \For {$m_i$ in measurements}
        \State $g_i \leftarrow \texttt{grid space where } m_i \texttt{ was taken}$
    \EndFor% \\
    \State \textbf{MEASUREMENT SELECTION}
    \For {$g_j$ in grid spaces}
        \State $sel\_meas_j \leftarrow \texttt{argmax}_{s.t. g_k = g_j} SNR_k$
    \EndFor %\\
    \vspace{-0.03in}
    \State \textbf{TRILATERATION}
    \State Perform Trilateration with all $sel\_meas$ 
\EndFor
\end{algorithmic}
\label{alg:selection}
\end{algorithm}

Table~\ref{table:isolation_nulling} reports the average cancellation for the vertical and horizontal pairs of antennas. It shows the cancellation for each frequency, and the average cancellation across frequencies. With an average cancellation of 23dB, \name\ can significantly mitigate the self-interference between parallel antennas. This is the cancellation on top of the natural isolation\cutt{ of the antennas}(due to the attenuation of the signal over the direct path). With a natural isolation of $\sim$20dB(achievable by either small antenna spacing or a circulator\cite{circulator}), the overall isolation is $>$40dB, roughly \cutt{equivalent}\textredd{equal} to that of \cutt{the }cross-polarized \cutt{in-band}\textredd{CP} signal\textredd{s}.

\vspace{-0.15in}
\section{Robust 3D Localization}\label{sec:3d}
\vspace{-0.03in}
So far, we have described how \name\ enables compact, efficient 1D localization for all tags regardless of orientation. As a user moves, the system collects 1D measurements across space to perform 3D localization via trilateration.

\begin{figure*}[t]
    \centering
    \begin{minipage}[t]{0.37\textwidth}
        \centering
        \includegraphics[width=0.85\linewidth]{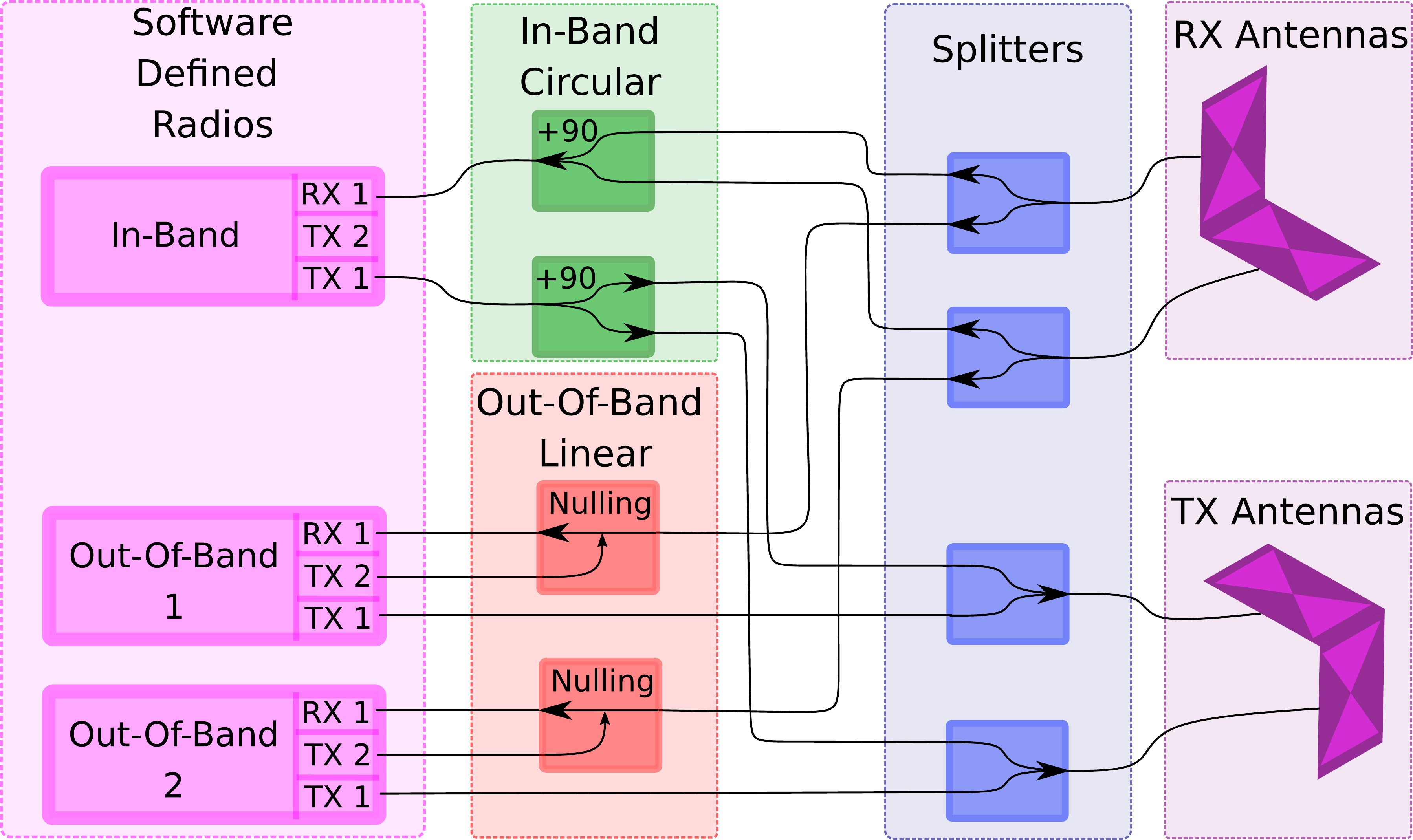}
        \caption{\footnotesize{Hardware Architecture.} \textnormal{The diagram shows SDRs, phase shifters, nulling, splitters, and antennas. }}
        \label{fig:schematic}
    \end{minipage}
    \centering
    \hspace{0.02in}
    \begin{minipage}[t]{0.2\textwidth}
        \centering
        \includegraphics[width=0.375\linewidth]{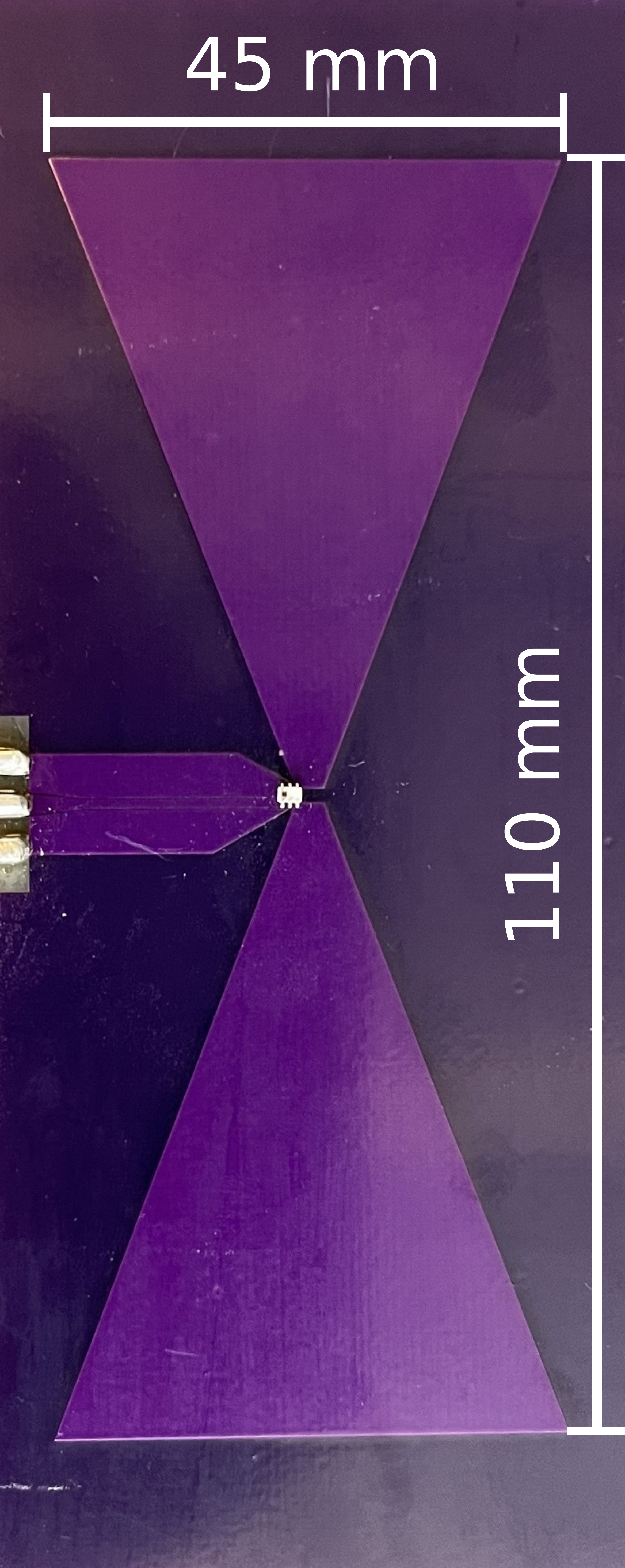}
        \caption{\footnotesize{Bowtie Antenna.} \textnormal{Our custom wideband antenna.}}
        \label{fig:antenna}
        \vspace{-0.2in}
    \end{minipage}
    \hspace{0.02in}
    \begin{minipage}[t]{.4\textwidth}
        \centering
        \hspace{-0.1in}
        \begin{subfigure}[t]{0.49\linewidth}
            \centering
            \includegraphics[width=1.095\linewidth]{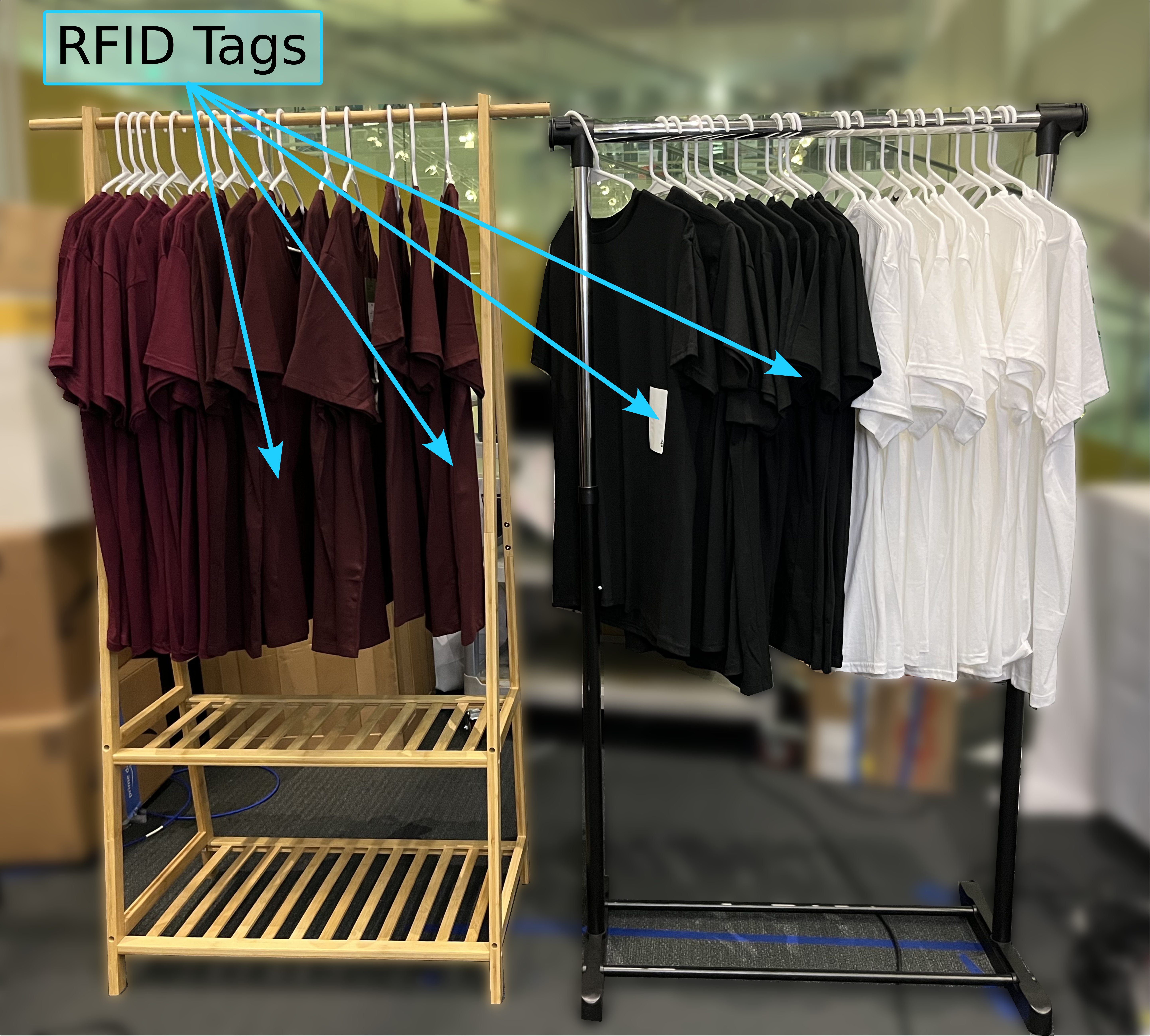}
        \end{subfigure}
        \hspace{0.07in}
        \begin{subfigure}[t]{0.42\linewidth}
            \centering
            \includegraphics[width=1.13\linewidth]{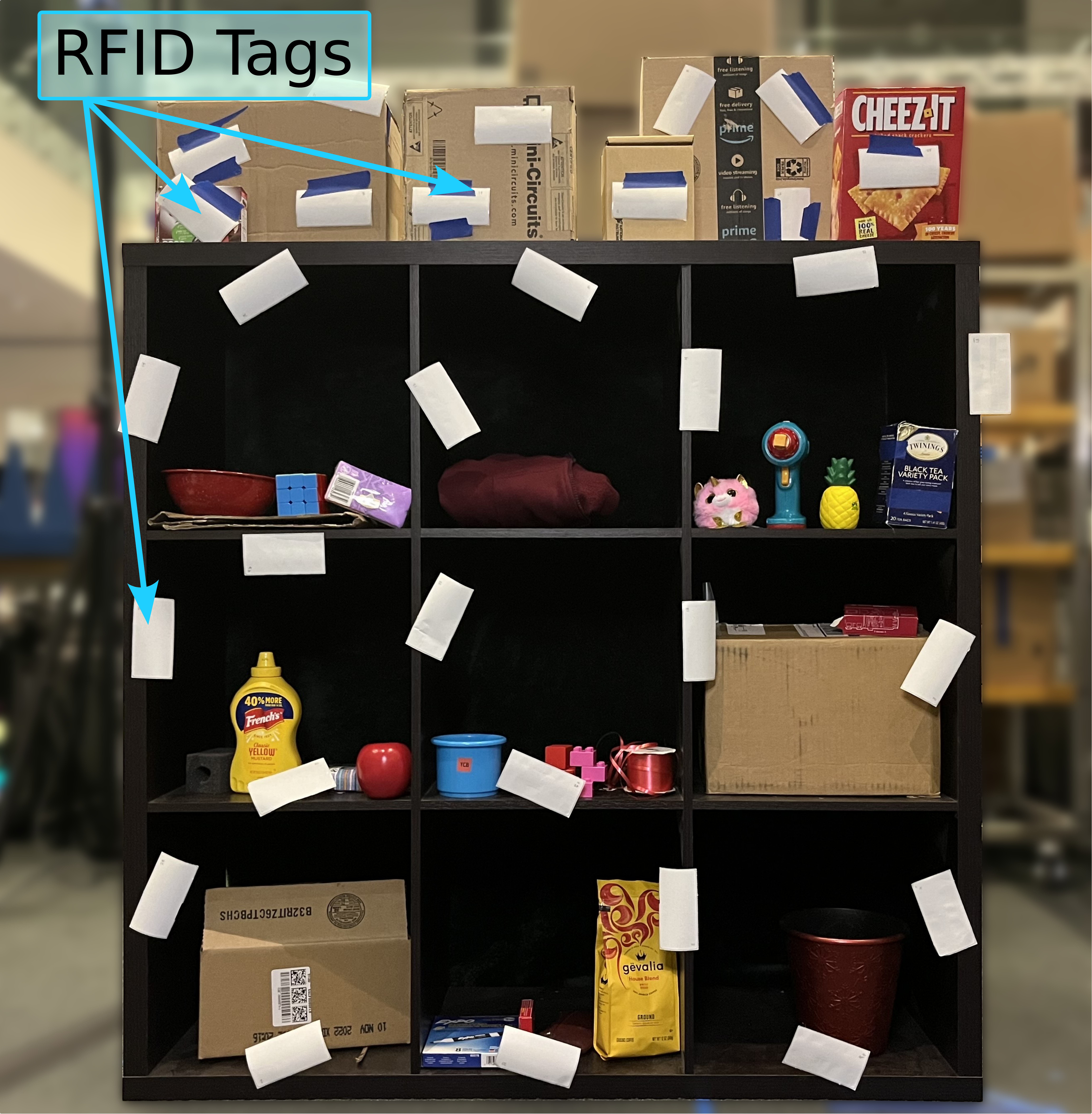}
        \end{subfigure}
        \caption{\footnotesize{Evaluation Environments.} \textnormal{This figure shows two example evaluation environments for \name.}}
        \label{fig:eval}
        \vspace{-0.1in}
    \end{minipage}
    \vspace{-0.15in}
\end{figure*}

One challenge in performing 3D localization is selecting more optimal vantage points for localization. Specifically, if the measurements are collected from nearby locations, their intersection will be sensitive to small errors in the 1D estimates. This is a well-known phenomenon in RF localization systems called dilution of precision (DoP). Ideally, the system needs to choose measurement locations that are furthest apart to reduce the probability of a poor DoP~\cite{RFusion}.

To do this, \name\ introduces an algorithm that intelligently selects a subset of its 1D measurements to use for trilateration. The goal of this selection is to maximize DoP and measurement SNR (in order to minimize  the likelihood of erroneous measurements for each tag and thus improve the robustness of localization). \name's measurement selection algorithm consists of the following three steps, which are detailed in Alg.~\ref{alg:selection}. The first step involves filtering to remove all measurements with an SNR below a threshold $\tau$ (e.g., 4~dB). This helps eliminate poor measurements that are likely to have high error. The second step involves sorting all measurements for a given tag based on their location in space and dividing the bounding box that contains them into $3\times3\times3$ evenly-spaced grid. The final step is selecting the measurement with the highest SNR from each grid.\footnote{If a grid space is empty, we select the measurement with the highest SNR.}

After selecting 1D measurements, \name\ performs trilateration with outlier rejection similar to~\cite{RFusion} to localize in 3D. This is repeated for every tag in the environment.

\vspace{-0.125in}
\section{Implementation \& Evaluation} \label{sec:implementation} \label{sec:evaluation}
\vspace{-0.03in}

\textbf{Hardware.} \name's hardware architecture is shown in Fig.~\ref{fig:schematic}. We implemented a wideband RFID reader following past designs~\cite{RFind,RFusion,RFChord}. We extended it to use three Nuand BladeRF software defined radio~\cite{bladeRF}: one for the CP signal, one for the horizontal LP sensing signal and one for the vertical LP. To create CP, we used two ZX10Q-2-13-S+ RF power splitters\cite{phase_splitter} to apply 90\textdegree\ phase shifts. 
At each antenna, the signals for the CP and LP were combined (or split) using  ZAPD-21-S+ splitters\cite{splitter}.

\noindent
\textbf{Custom Antennas.} Various antenna designs have been proposed for compact UHF RFID readers~\cite{PolDivAntenna,SlantPol, compactwideband,compactminiaturized}. We custom-designed our antennas to have a small factor and desired frequency range (700~MHz to 1~GHz). We selected a bowtie design (Fig.~\ref{fig:antenna}) due to its ability to achieve relatively flat wideband operation in small form factor. 
Our antenna was fabricated on a 0.8mm-thick FR4 substrate and measures 4.5cm x 11cm. Since the bowtie antenna is a balanced structure, it was necessary to add a balun (balanced to unbalanced component) between its two branches to efficiently connect it to a coaxial cable (an unbalanced structure). 

\noindent
\textbf{Software.} We connected the BladeRFs to a raspberry pi\cite{rpi} to collect RFID measurements. \textred{We implemented the EPC Gen 2 Protocol~\cite{epcGen} for reading RFID tags, allowing us to read any number of tags in the environment. } Self-localization was implemented using the Intel Realsense T265 camera\cite{camera}, which has\cut{ a} built-in visual-inertial odometry (VIO). We synchronized the output of the camera with the samples obtained from the BladeRFs. We processed the measurements and computed 3D location estimates on an Ubuntu 20.04 computer. We used the SciPy\cite{scipy} library to perform trilateration. 

\noindent \textbf{\textred{Latency.}} \textred{In \name, self-interference cancellation and RFID interrogation are performed in real-time. In our implementation, a tag can be read in approximately 6ms.\footnote{\textred{Faster tag reading can be achieved by changing the backscatter link frequency as part of the EPC Gen 2 protocol\cite{epcGen}.}} Our processing is currently performed in post-processing on a separate computer, but can be run in parallel with RFID data acquisition on the edge (e.g., raspberry pi) or in the cloud to achieve real-time operation. Currently, a ToF estimate takes an average of 13ms to compute, and the end-to-end localization (Alg.~\ref{alg:selection}) can be run on separate threads in roughly 750ms. }

\noindent
\textbf{\textred{Transmission Frequencies.}} 
\textred{As described in \xref{sec:jtdl}, we transmit both a CP signal to power the tags and LP signals to localize the tags. We can transmit our CP signal inside the UHF ISM band at around 30~dBm. For our LP signals, we transmit with an average EIRP less than -20dBm to meet FCC regulations. To choose the transmission frequencies, we selected multiple frequencies to span the roughly 250~MHz of bandwidth that COTS RFIDs reflect\cite{RFind}. Given the PLL relock time of our SDR (i.e., time to switch frequencies), we could fit 8 frequency hops within each tag read. Choosing near-equal spacing between frequencies, while avoiding high interference frequencies (including in-band interference),  lead us to the frequencies listed in Table~\ref{table:isolation_nulling}. \footnote{\textred{Given our out-of-band frequencies, our maximum unaliased range (due to subsampling in the frequency domain) is roughly 5m. While this is sufficient for our implementation, this could be extended by changing the frequencies to be closer together.}} }

\begin{figure*}[t]
    \centering
    
    \begin{minipage}[t]{.32\textwidth}
        \centering
        \includegraphics[width=\linewidth]{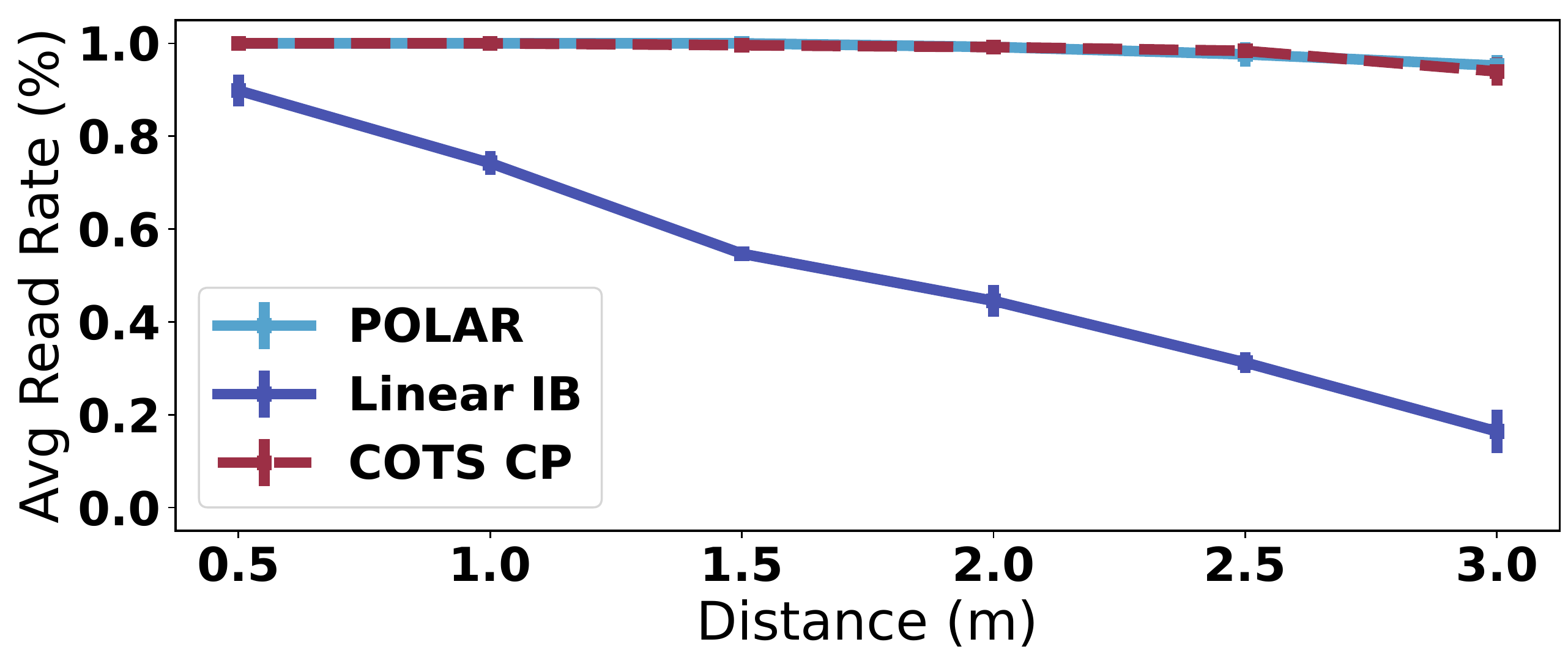}
        \vspace{-0.2in}
        \caption{\footnotesize{Read Rate vs Distance.} \textnormal{Average read rate of \name\ (blue),\cut{ and} \emph{Linear IB} (purple)\textred{, and \emph{COTS CP}(red)} vs. distance\cutt{ from the plane of reader measurements to the plane of the tags}.}}
        \label{fig:mb_read_rate_0_90}
    \end{minipage}
    \hspace{0.01\linewidth}
    \begin{minipage}[t]{.32\textwidth}
        \centering
        \includegraphics[width=\linewidth]{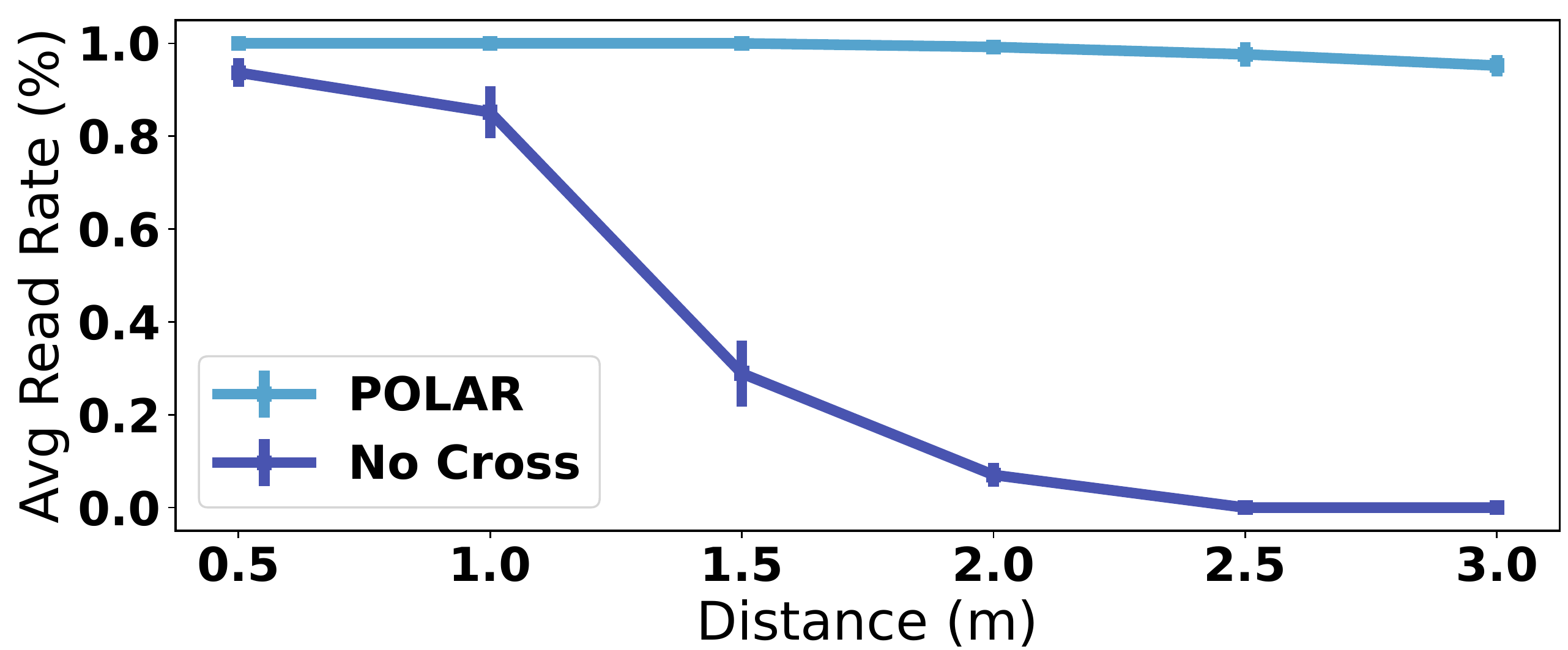}
        \vspace{-0.2in}
        \caption{\footnotesize{Benefit of Cross-Polarization.} \textnormal{Average read rate vs. distance for \name\ (blue) and a system without cross-polarization (purple).}}
        \label{fig:mb_no_cross}
    \end{minipage}
    \hspace{0.01\linewidth}
    \begin{minipage}[t]{0.32\textwidth}
        \centering
        \includegraphics[width=\linewidth]{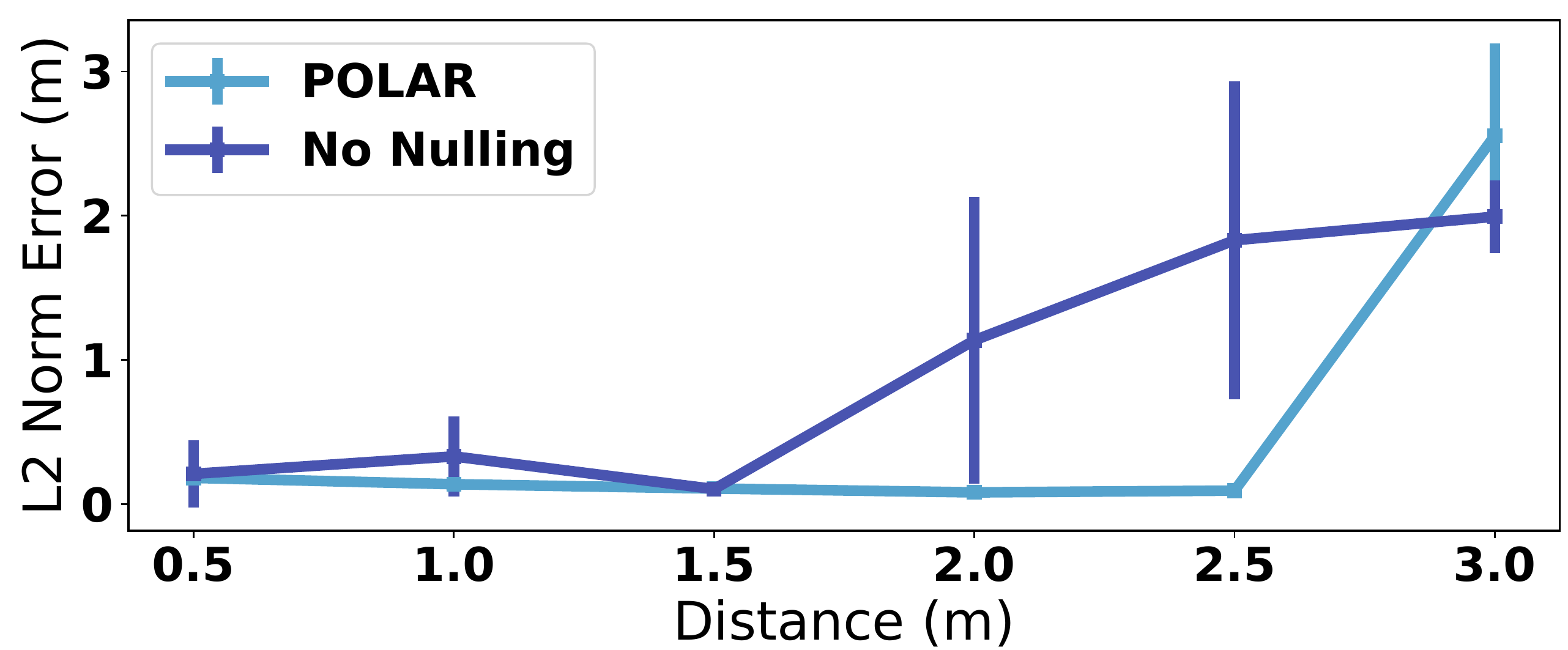}
        \vspace{-0.2in}
        \caption{\footnotesize{Benefit of Over-the-Wire Nulling.} \textnormal{Average 2D localization error vs. distance for \name(blue) and \emph{No Nulling}(purple).}}
        \label{fig:mb_no_nulling}
    \end{minipage}
    \vspace{-0.15in}
\end{figure*}

\noindent
\textbf{Evaluation Environment.} We evaluated \name\ in a standard indoor environment in an office with tables, chairs, computers, etc. In our experiments, there were users walking in the background and other standard wireless technologies. \textred{Unless stated otherwise, our \name\ device was moved through the environment by a human user. }

Our evaluation covered two types of settings, as shown in Fig.~\ref{fig:eval}. The first involved standard RFID-tagged items in the environment (e.g., clothes). The second involved placing multiple RFID tags at various angles on a wooden shelf filled with everyday objects. This setup allowed us to perform benchmark testing at different angles and distances, to study the impact of orientation. 
We used standard off-the-shelf RFID tags such as the Alien Squiggle RFIDs\cite{rfid}.

%\vspace{0.05in}
\noindent
\textbf{Ground-Truth.} Ground truth was obtained from a motion capture system (OptiTrack~\cite{optitrack}), with ceiling-mounted infrared cameras and infrared markers placed on  target items. 

\vspace{-0.125in}
\section{Microbenchmarks}
\vspace{-0.03in}

\textred{We performed microbenchmark experiments to evaluate the impact of various factors on \name's performance.}

% \textred{We performed microbenchmark experiments to evaluate the impact of various factors on \name's performance. \footnote{\textred{Additional microbenchmarks are available in the extended version\cite{extended_polar}.}}}

\vspace{-0.1in}
\subsection{Read Range in CCP vs LP}
\vspace{-0.03in}

In our first microbenchmark, we evaluated how \name's CCP design improves its ability to read tags at longer distances. 

\noindent
\textbf{Baseline:} We compared \textred{to two}\cut{a} partial implementation\textred{s.}\cut{,} \textred{The first implementation, }\emph{Linear IB,}\cut{that} did not generate circularly polarized signals for in-band. Instead, it used a single pair of horizontally polarized antennas (transmit and receive). Since the baseline cannot leverage cross-polarization to limit self-interference, we implemented over-the-wire nulling(described in~\xref{sec:nulling}), allowing this implementation to transmit the same power as \name. \textred{The second implementation, \emph{COTS CP}, used a commercial CP antenna\cite{PatchAntenna} to generate circularly polarized signals. We calibrated the transmit power of this implementation such that it had the same equivalent isotropic radiated power (EIRP) as \name. }

\noindent
\textbf{Experiment:}
We placed over 30 tags at various angles in a plane, with a roughly even distribution of angles. To quantify the impact of distance on read rate, we moved the antennas in a plane parallel to the one with the tags. We repeated the same movement pattern for the baselines. Across trials, we varied the distance from the plane to the tag. In each trial, we measured the read rate as the percentage of tags read.

\noindent
\textbf{Result:}
Fig.~\ref{fig:mb_read_rate_0_90} plots the average read rate as a function of distance for \name\ (blue), \emph{Linear IB} (purple)\textred{, and \emph{COTS CP} (red - dashed)}. The error bars show standard deviation. We make the following remarks: 
\vspace{-0.03in}
\begin{itemize}
    \item At the closest distance of 0.5~m, \name\ reads 100\% of tags while \emph{Linear IB} reads 90\% of tags. This relatively high percentage for \emph{Linear IB} is due to the fact that at close distances, even tags with large polarization mismatches are able to harvest enough energy to power. Only tags very close to perpendicular are unable to be read. However, \name\ is still able to read 10\% more tags than the baseline. Also, in practical environments it is infeasible to be within 0.5~m of all tags, as this would require a very large scan of the environment, defeating the benefits of a handheld RFID reader. 
    \item As the distance increases, \name's performance remains consistent while \emph{Linear IB}'s performance drops drastically. For example, at 3~m, \name\ is still able to read 95\% of tags. In contrast, \emph{Linear IB} is only able to read 16\% of tags. \textred{This demonstrates the importance of \name's CCP design for generating CP signals to read all tags in the environment, even at long ranges.}
    \item \textred{Across all distances, the performance of \name\ and \emph{COTS CP} match very closely (e.g., both 99\% at 2~m and 95\% vs 94\% at 3~m). This shows that \name's CCP design is able to successfully create CP signals to power tags across all angles, matching the performance of a commercial antenna. }
\end{itemize}
\vspace{-0.075in}

\vspace{-0.2in}
\subsection{Impact of Cross-Polarization on Range}
Recall from~\xref{sec:isolation} that RFID readers perform self-interference cancellation for full-duplex operation. \name\ adopts a cross-polarization approach for the in-band CP transmission. In our next microbenchmark, we evaluated the improvement that this provides in the read range of the device. 

\begin{figure*}[t]
    \centering
    \begin{minipage}[t]{0.32\textwidth}
        \centering
        \includegraphics[width=\linewidth]{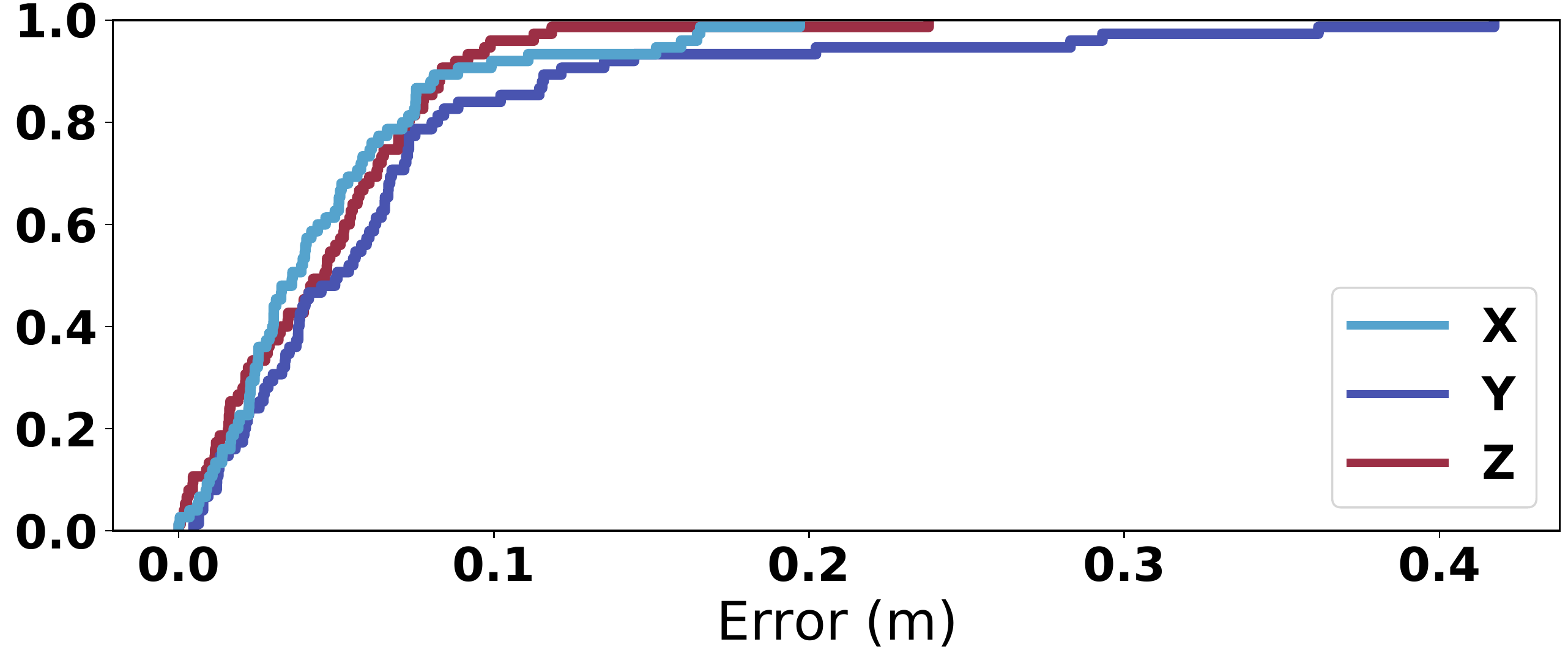}
        \vspace{-0.2in}
        \caption{\footnotesize{Localization Accuracy.} \textnormal{CDF of \name's error in X(blue), Y(purple), and Z(red).  }}
        \label{fig:localization_accuracy_xyz}
    \end{minipage}
    \hspace{0.01\linewidth}
    \begin{minipage}[t]{.32\textwidth}
        \centering
        \includegraphics[width=\linewidth]{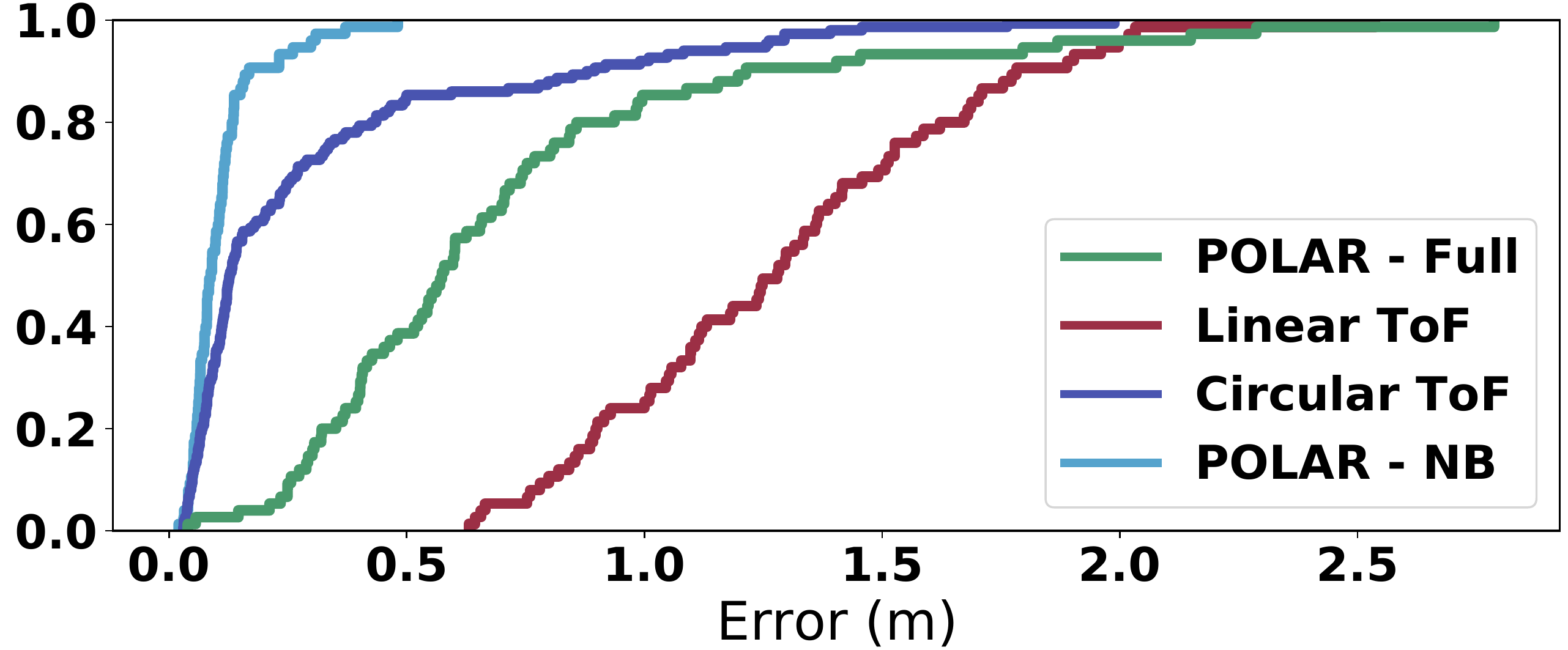}
        \vspace{-0.2in}
        \caption{\footnotesize{\cut{Localization Accuracy.}\textred{Baseline Comparison.}} \textnormal{CDF of L2 norm error for \name(blue),\cut{ and} \emph{Linear \textred{ToF}}(purple)\textred{, \emph{Circular \textred{ToF}(red)}, and \name-Narrowband(green)}.  }}
        \label{fig:localization_accuracy_baseline}
    \end{minipage}
    \hspace{0.01\linewidth}
    \begin{minipage}[t]{.32\textwidth}
        \centering        
        \includegraphics[width=\linewidth]{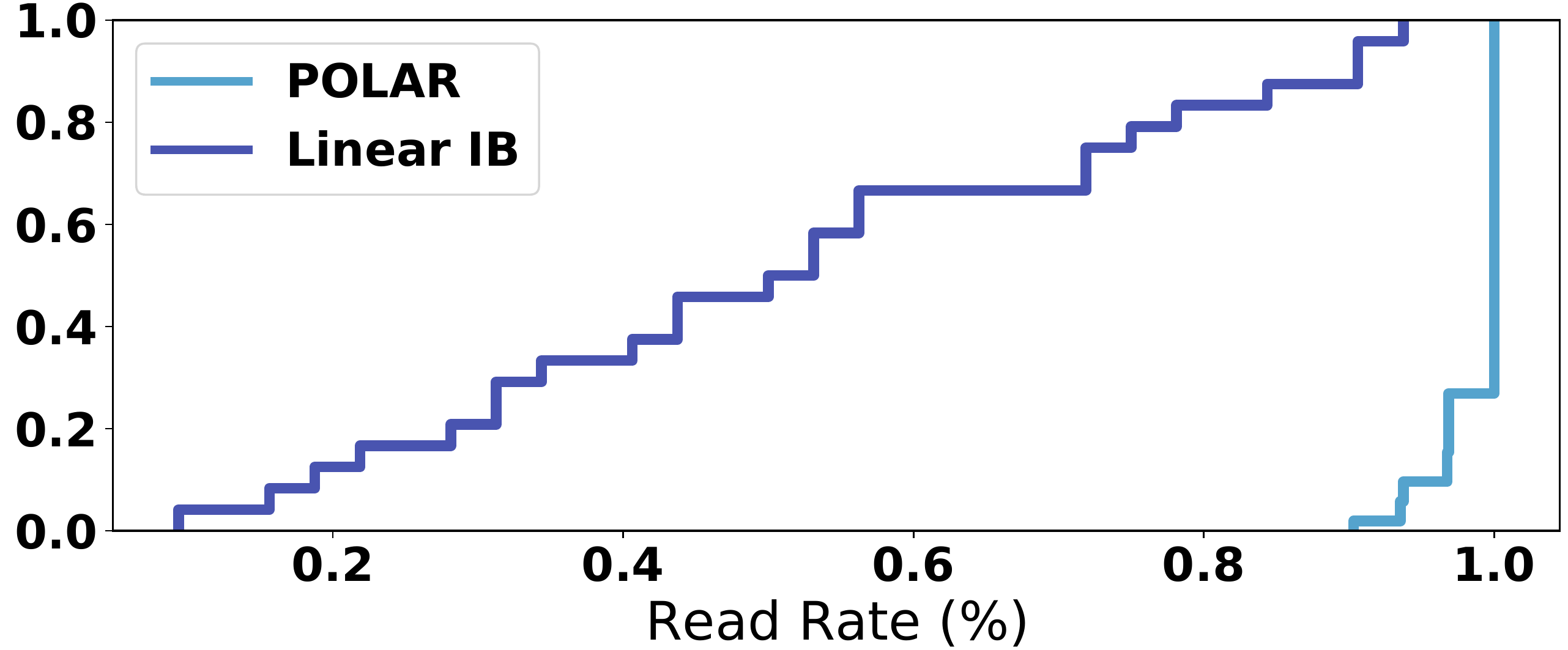}
        \vspace{-0.2in}
        \caption{\footnotesize{\cut{Localization Accuracy.}\textred{Read Rate.}} \textnormal{CDF of the read rate for \name\ (blue), \emph{Linear IB} (purple).  }}
        \label{fig:ib_read_rate}
    \end{minipage}
    \vspace{-0.15in}
\end{figure*}

\noindent
\textbf{Baselines:} We compared to a partial implementation of our system without cross-polarization (i.e., it transmits RHCP and receives LHCP signals). Due to the higher level of self-interference in this implementation, we had to decrease the transmit and receive gains to prevent clipping.

\noindent
\textbf{Experiment:}
We placed over 30 tags in a plane in the environment at various angles ranging from 0\textdegree\ to 360\textdegree. We moved our reader in a plane parallel to the tags and used it to read the tags. We repeated this at different distances, each time measuring the read rate (percentage of tags in the environment that were read) for a given distance. 

\noindent
\textbf{Result:}
Fig.~\ref{fig:mb_no_cross} plots the average read rate as a function of distance for \name\ (blue), and the system without cross-polarization (purple). The error bars denote the standard deviations. 
We observe that \name\ consistently reads tags at distances up to 3~m, with the read rate only dropping from 100\% to 95\% over this range. In contrast, without cross-polarization, the read rate drops significantly with distance. For example, it is only able to read 29\% of tags at 1.5m, and cannot read any tags at 2.5m and beyond. This shows the value of our cross-polarization for reading tags at long ranges. 

\vspace{-0.1in}
\subsection{Impact of Nulling on Range} \label{sec:nulling_range}
\vspace{-0.03in}
Next, we evaluated the benefit of \name's nulling, which is used to cancel LP self-interference (per~\xref{sec:nulling}).

\noindent
\textbf{Baseline:} We compared to a partial implementation without nulling. Due to the higher level of self-interference, we had to decrease the transmit and receive gain to avoid clipping.

\noindent
\textbf{Experiment:}
We used a similar setup to the one described above, this time placing six tags at various angles in a plane. To quantify the impact of distance on localization accuracy, we moved \name's antennas in a plane parallel to, and at a fixed distance from, the one containing the tags; across experimental trials, we changed the plane's distance to the tags. Since \name's movement was constrained to a 2D plane, we only focused on 2D localization here. We computed the error as the L2-norm between the ground truth tag location and the estimated tag location.

\noindent
\textbf{Result:}
Fig.~\ref{fig:mb_no_nulling} plots the average 2D error as a function of distance for \name\ (blue) and the partial implementation (purple). The error bars denote the standard deviation. 

We observe that \name\ accurately locates tags up to 2.5m  with only 9cm error. Beyond 3m, the error increases with distance as we cannot transmit higher power due to FCC regulations\cite{RFind}. In contrast, without nulling, the system cannot accurately localize past 1.5m, with an error of 1.1m at 2m. This shows the value of nulling in \name's design.

\vspace{-0.1in}
\section{Performance Results}

\vspace{-0.03in}
\subsection{Localization Accuracy}

\vspace{-0.03in}
To evaluate \name's overall localization accuracy, we conducted an experiment where different RFID tags were placed in random locations and orientations (ranging from 0\textdegree\ to 360\textdegree) in the  fully-furnished experimental environment described in~\xref{sec:evaluation}. \textred{Unlike the previous microbenchmarks, these tags were not contained to a single plane.} Each trial included at least 25 RFIDs. A user moved a \name\ device around the environment. During each trial, the system self-localizes using computer vision, estimates the time-of-flight to different tags, and then combines the estimates over space to estimate each tag's 3D location. We captured over 75 location measurements in total. Each RFIDs location was accurately measured to obtain the ground truth, and the error is computed as the difference between the estimated location and the ground truth location along each of the X/Y/Z dimensions. 

Fig.~\ref{fig:localization_accuracy_xyz} plots the localization error CDF in the X (blue), Y (purple), and Z (red) dimensions. We note the following:
\vspace{-0.03in}
\begin{itemize}
    \item \name\ localized RFIDs with a median accuracy of \cut{5}4~cm, 5~cm, and \cut{7}5~cm in the X, Y, and Z dimensions respectively. This shows \name\ can achieve very high (sub-10~cm) median accuracy in challenging indoor environments.
    \item \name's 90\textsuperscript{th} percentile is \cut{14}9~cm, \cut{11}12~cm, and \cut{14}8~cm in the X, Y, and Z dimensions, respectively. This shows that \name's localization is robust to location and orientation.
\end{itemize}

\vspace{-0.03in}

\noindent \textbf{\cut{Comparison to a Partial Implementation:}\textred{Baseline Comparisons:}}
Next, we compared to \cut{a partial implementation, \emph{Linear}, which relies on only one linearly polarized antenna pair (transmit and receive) for localizing. We evaluated this using both horizontal and vertical antenna pairs.}\textred{two baselines, where we modified state-of-the-art RFID localization\cite{RFind} to operate in a portable form with a fixed polarization. For the first baseline, \emph{Linear ToF}, we transmitted fixed linear polarizations (either horizontal or vertical) to localize the tags.} To isolate the impact of powering up on localization, \cut{our partial implementation}\textred{\emph{Linear ToF}} incorporated \name's approach for CCP-based powering (by generating circularly polarized signals) but not its approach for localization. Furthermore, to ensure fair comparisons, we used the same measurements obtained from the above experiment, the main difference being that in \emph{Linear ToF}, we only provided the measurements from one pair of antennas (either horizontal or vertical) to the localization algorithm. \textred{For our second baseline, \emph{Circular ToF}, we transmitted circularly polarized signals to localize the tags. We used COTS CP antennas\cite{PatchAntenna} to generate both the CP powering and localization signals, and we leveraged the cross-polarization technique from \xref{sec:circcancel} to limit the self-interference of both the powering and localization signals. We calibrated this baseline to transmit the same equivalent isotropic radiated power (EIRP) as \name. Furthermore, for this baseline, we provided ground-truth antenna locations (from the Optitrack motion capture system\cite{optitrack}) instead of those derived from VIO self-tracking. }  Aside from CCP-based polarization projection, \cut{the implementation applies}\textred{both our baselines apply} the same algorithms for 3D localization. The localization error is computed as the L2-norm between the ground truth location and the estimated location.

Fig.~\ref{fig:localization_accuracy_baseline} plots the CDF of the localization error \cut{across trials }for \name\ (blue), \cut{and }\emph{Linear ToF}(purple)\textred{, and \emph{Circular ToF} (red)}. \cut{Note the following:}\textred{We note:}

\vspace{-0.03in}
\begin{itemize}
    \item \name's median and 90\textsuperscript{th} percentile accuracy are \cut{12}9~cm and \cut{24}17~cm respectively. Note that the L2 norm here matches what one would expect from the earlier reported numbers in the X/Y/Z dimensions, and demonstrates \name's high 3D localization accuracy
    \item \emph{Linear ToF} has a median accuracy of \cut{16}12~cm. High median accuracy is expected since \emph{Linear ToF} obtains good SNR across more than half the orientations as per our investigation of Fig.~\ref{fig:ib_snr_ours}. \footnote{\textred{We do note that compared to \emph{Linear ToF}'s performance, higher RFID localization accuracy has been reported in literature. However, past work that achieved higher accuracy either used large antenna arrays\cite{AoA, PhasedArray, TwodimensionLO} or robots\cite{Fusebot, RFusion, MobiTag, PinIt} moving on predefined trajectories, neither of which are suitable for handheld human moblility.}}
    \item However, \emph{Linear ToF} has much poorer 90\textsuperscript{th} and 95\textsuperscript{th} percentile accuracy than \name's full implementation. Specifically, it achieves a 90\textsuperscript{th} percentile of \cut{82}88~cm, and its 95\textsuperscript{th} percentile is \cut{1.05}1.22~m (\name's is \cut{$29~cm$}27~cm). These errors (>3 times that of \name) are due to the polarization mismatch between the antennas and some of the tags, causing low SNRs. This shows the importance of \name's design to \textit{accurately} and \textit{robustly} localize tags across all angles.
    \item \textred{Finally, \emph{Circular ToF} achieves a much worse median (1.28~m) and 90\textsuperscript{th} percentile (1.78~m) accuracy. This is expected, since the circular polarization leads to unknown phase offsets due to tag orientations, causing large localization errors. }
\end{itemize}

\vspace{-0.03in}

\noindent \textbf{\textred{Impact of UWB:}} \textred{Recall from \xref{sec:jtdl} that \name\ leverages its JTDL design in order to measure UWB RFID channel measurements in real-time. To demonstrate the benefit of UWB measurements for localization, we compared the performance of \name\ to a partial implementation (\name- Narrowband) that does not rely on UWB measurements. The partial implementation relies on 43MHz of bandwidth, instead of \name's 245MHz of bandwidth. To ensure a fair comparison, both systems rely on the same measurements, except the partial implementation only leverages a limited bandwidth for each measurement. }

\textred{Fig.~\ref{fig:localization_accuracy_baseline} plots the CDF of the localization accuracy for \name\ (blue) and \name-Narrowband (green).}
\textred{We observe that \name\ significantly outperforms \name-Narrowband, achieving more than a 6 times improvement in both the median (9cm vs. 57cm, respectively) and  the 90\textsuperscript{th} percentile (17~cm vs. 121~cm, respectively). This level of improvement is expected, since narrowband measurements are unable to accurately localize RFID tags in the presence of multipath (as demonstrated in \cite{RFind}). This result demonstrates the importance of \name's JTDL design to enable these UWB measurements in real-time. }

\vspace{-0.1in}
\subsection{Read Rate}
\vspace{-0.03in}

Finally, we evaluated \name's ability to successfully read all tags in the environment. Recall from~\xref{sec:circ} that \name\ generates a circularly polarized signal to read tags regardless of orientation. We compare its performance to a baseline: \emph{Linear IB}. In this implementation, we use a horizontally polarized antenna (as opposed to \name's circularly polarized signal) to power and read the tags. The baseline uses nulling for self-interference cancellation (similar to ~\xref{sec:nulling}), allowing it to transmit the same power as \name.

We ran 76 experimental trials, each with more than 30 tags in the environment at various angles (from 0\textdegree\ to 360\textdegree). The user moved a \name\ device around the environment, and the device continuously queried the tags. We moved the device along the same trajectories for both systems. 

Fig.~\ref{fig:ib_read_rate} plots a CDF of the read rate (the percentage of tags read) for \name\ (blue) and \emph{Linear IB} (purple). We note:

\vspace{-0.03in}
\begin{itemize}
    \item \name\ reads over 90\% of the tags in all experiments, and can read 100\% of the tags in over 70\% of the experiments. 
    \item In contrast, \emph{Linear IB} has a median read rate of 52\%. This is due to polarization mismatch between the antenna and some tags, preventing those tags from powering (as in Fig.~\ref{fig:ib_snr_ours}). 
    \item Interestingly, in some trials, the baseline can read up to 94\% of tags. This is because at very close distances (<0.5m), even tags with a large polarization mismatch are able to harvest enough energy to power. 
    However, at longer distances, polarization mismatch leads to poor performance. 
\end{itemize}

\vspace{-0.03in}
This shows the importance of our CCP design for sending CP signals to power tags across all orientations. 
\vspace{-0.1in}
\section{Related Work} \label{sec:related}
\vspace{-0.03in}
\noindent
\textbf{Fine-Grained RFID Localization.} 
Early research on RFID localization used received signal strength (RSS)~\cite{T21,LANDMARC} (which suffers in typical indoor environments due to multipath). Later advances employed more sophisticated techniques (antenna arrays~\cite{AoA,PhasedArray,TwodimensionLO} and time-of-flight~\cite{RFind,TurboTrack,RFChord,RFGrasp}) to achieve high-localization accuracy. However, state-of-the-art techniques typically require bulky setups with antennas separated by meter-length distances, making it infeasible for handheld devices. To avoid bulky infrastructures, researchers have considered mounting antennas on mobile robots to emulate antenna arrays (i.e., SAR)~\cite{PinIt,RFusion,MobiTag,Fusebot}; however, these methods require moving the robot and/or tag on well-defined trajectories (e.g., on a track at constant velocity), making them ill-suited for natural human mobility. \name\ builds on this literature, introducing new mechanisms to bring fine-grained localization to handheld readers.

\noindent
\textbf{Handheld Readers.} 
The vast majority of existing handheld RFID readers (e.g., Zebra\cite{ZebraHandheld}, Bluebird\cite{BluebirdHandheld}, and AsReader\cite{AsReaderHandheld}) can read tags but not localize them. These devices employ a single circularly polarized antenna, limiting their ability to perform phase-based localization (see~\xref{sec:problem}). Instead, prior work proposed deploying a dense surveyed grid of reference tags and using fingerprinting-based localization\cite{gimpilevich2015design}; however, this has high overhead and limited scalability. 

To localize tags, existing handheld readers implement a ``geiger" counter mode which repeatedly queries a target tag to measure its RSS. As a user moves, the device beeps louder when the RSS increases to help the user locate the tag~\cite{GeigerCounter}. This approach is time-consuming and inefficient as it searches for a single tag at a time. Recent research extended this idea by also leveraging phase~\cite{chatzistefanou2022target,singh2020localization}; these proposals require a user to move in specific patterns (e.g., make a large circle in the air) then apply antenna-array equations on the trajectory. Thus, similar to the geiger approach, they are time-consuming and not scalable for typical environments with hundreds or thousands of tags. \name\ shares the vision of these proposals and introduces new advances (CCP and JTDL) to enable accurate localization from each tag response, making its design scalable, portable, and efficient. 

\noindent
\textbf{Orientation in RFID Localization.}
Tag orientation is a known problem in RFID localization systems. There have been two main approaches to deal with it, but neither is suitable for handhelds. The first relies on bulky antenna arrays (or SAR on predefined trajectories), whose formulation is independent of orientation (since it uses phase differences between antennas)~\cite{PinIt,Fusebot,MobiTag,reloc}. Variants of these designs use a smaller number of antennas to factor out the phase, but are limited to line-of-sight conditions and cannot operate in practical multipath-rich environments~\cite{selby2018rapid,OmniTrack_2021,RFPrism}.
The second category of systems places multiple RFID tags on a target object to detect its orientation~\cite{Tagyro,RFCompass,MultiTagAngle1,MultiTagAngle2}, but adding multiple tags increases cost per-item. Other work requires specialized tags with IMUs~\cite{IMUTag} to detect orientation. Thus, we still lack a handheld solution that deals with orientation in the prevalent scenario of single tag per item.

\name\ is also related to prior work that measures RSS on two different antennas (vertically and horizontally polarized) to detect tag angle~\cite{PolarDraw,OrientationDetection,OrientationDetection2}. While one could in principle implement these in a handheld, their reliance on RSS for \textit{localization} limits them to line-of-sight conditions and makes them vulnerable in typical multipath-rich environments. We are inspired by this and extends it to enable accurate localization in standard indoor environments.

Finally, we build on prior work on self-interference cancellation~\cite{CrossPol1,lasser2015self,bharadia2013full} and dual-frequency excitation~\cite{RFind, RFusion, RFGrasp,TurboTrack,RFChord}. These techniques have been used in many prior systems and we customize them in our overall design.

\noindent
\textred{\textbf{Antenna Design for RFID readers.} \name\ is related to past work that aims to increase read range of RFID readers by specialized antenna designs including dual-polarized~\cite{PolDivAntenna}, slant-polarized~\cite{SlantPol}, compact~\cite{compactwideband}, circularly polarized~\cite{compactminiaturized}, etc. It is also related to past work on antenna design in general that switches between polarizations using methods such as phase shifters\cite{antenna_phase_shift},  PIN diodes\cite{antenna_pin_diodes, antenna_pin_diodes_2}, and varactor diodes\cite{antenna_varactor_diode}. \name\ builds on this prior literature and extends it in two key ways. First, rather than being limited to fixed polarization or switching between different polarizations~\cite{SwitchPol1,SwitchPol2}, \name\ can construct arbitrary linear and circular polarizations via CCP; second, our system goes beyond antenna control to building an end-to-end portable system for RFID discovery, identification, and localization.} 

\vspace{-0.1in}
\section{Discussion}
\vspace{-0.03in}
\noindent \textbf{Power Consumption.} \textred{While our prototype of \name\ relies on software-defined radios and COTS components, a production system would rely on a custom designed, fully-integrated PCB with RFID reader hardware, out-of-band transmitters/receivers, and VIO tracking that would be designed for optimal power consumption. Since out-of-band transmit power is limited by FCC regulations (as described in~\xref{sec:implementation}), we expect an optimized system to achieve similar battery life to existing handheld RFID readers\cite{ZebraHandheld, BluebirdHandheld}.}

\noindent \textbf{Range.} \textred{Recall from~\xref{sec:nulling_range} that at 3~m, \name\ is not able to successfully locate RFID tags due to low SNR. While this range is more limited than some infrastructure-based RFID systems\cite{ImpinjXarray}, it is on par with existing handheld RFID readers, which typically operate at 3-5~m (to conserve battery power by not transmitting higher power). Therefore, \name\ can be used in similar scenarios as existing handhelds, while providing fine-grained localization in addition to simply reading the tags in the environment. In future work, it would be interesting to extend \name's range even further. }

\vspace{-0.1in}

\section{Conclusion}
We presented a handheld system for fine-grained RFID localization and demonstrated it in practical indoor environments. Core to our design is a new approach for software-based polarization (CCP) which enables generating arbitrary linear and circular polarizations to optimally read and localize tags across orientations. While this approach was demonstrated in the context of handheld readers, its benefits extend to stationary readers, where it can increase read rate and tag SNR over prior designs. The paper's contributions go beyond developing CCP to building an end-to-end handheld localization prototype. By bringing fine-grained localization to handheld form factors, this work charts a more cost-effective and scalable way to its adoption.

\vspace{0.05in}
\noindent\footnotesize{\textbf{Acknowledgments} 
We thank the anonymous reviewers and the Signal Kinetics group for their help and feedback. This research is sponsored by NSF (Awards \#1844280 and \#2044711), the Sloan Research Fellowship, and the MIT Media Lab.}

\bibliographystyle{ACM-Reference-Format}
\bibliography{ourbib2}

\clearpage
\appendix

\normalsize

\section*{Appendix}

\section{Phase vs Orientation}

\begin{figure}[t]
    \centering
    \begin{subfigure}[t]{0.48\linewidth}
        \centering
        \includegraphics[width=\linewidth]{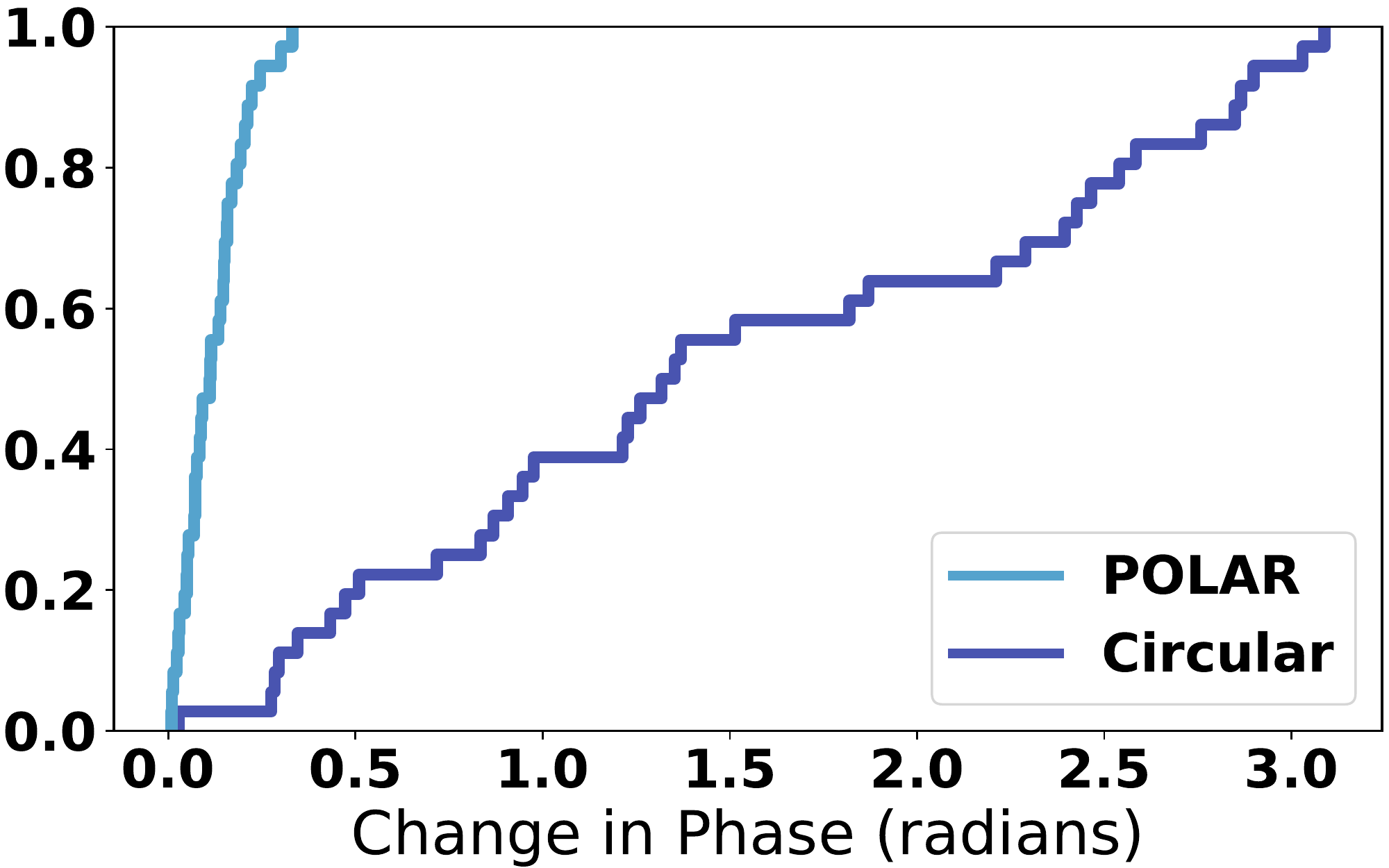}
        \vspace{-0.2in}
        \caption{\footnotesize{Change in Phase.}}
        \label{fig:mb_phase}
    \end{subfigure}
    \hspace{0.01in}
    \begin{subfigure}[t]{0.48\linewidth}
        \centering
        \includegraphics[width=\linewidth]{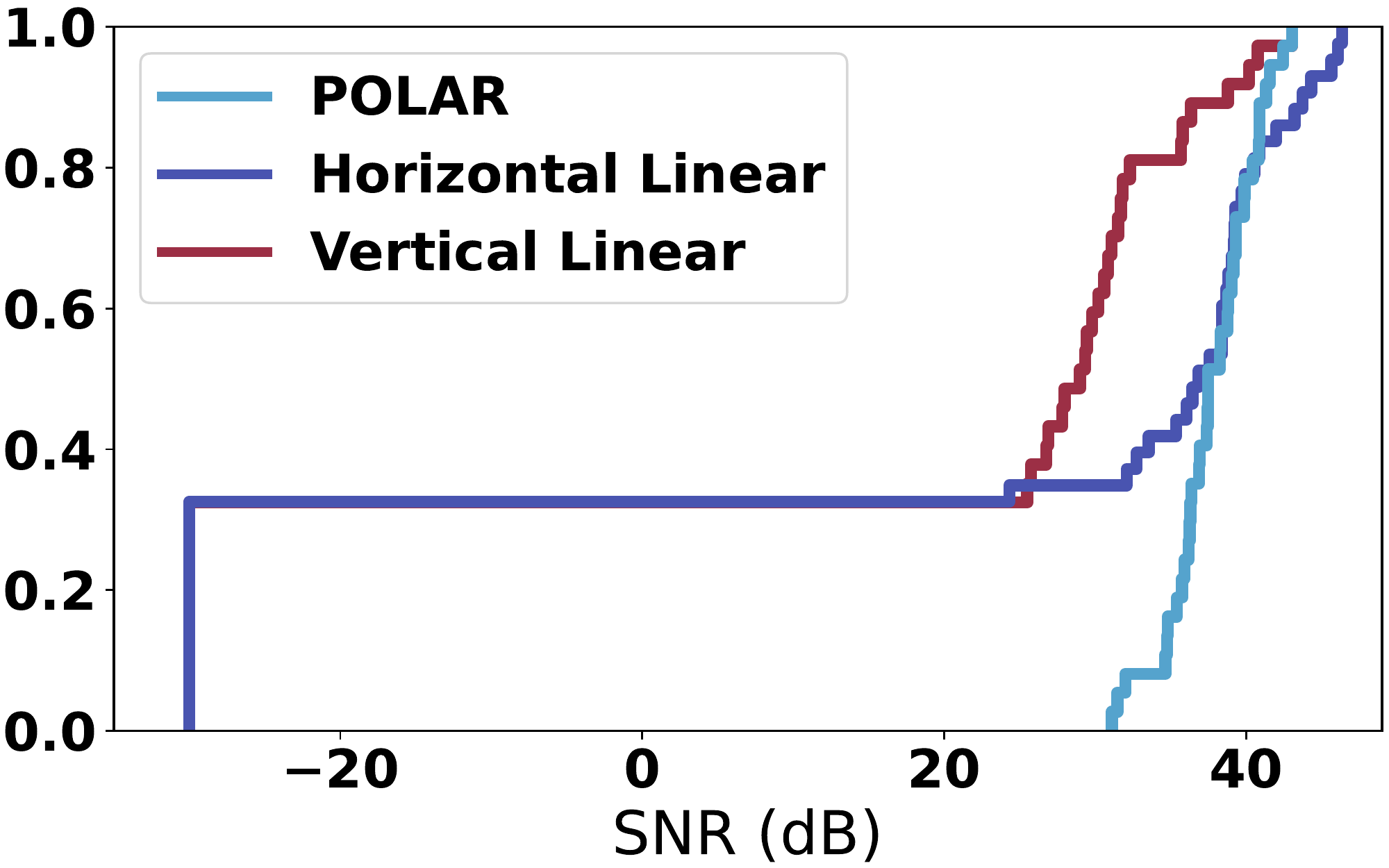}
        \vspace{-0.2in}
        \caption{\footnotesize{SNR of Response.}}
        \label{fig:mb_snr}
    \end{subfigure}
    \caption{Phase \& SNR across roll rotations. \textnormal{(a) CDF of the change in phase for \name\ (blue) and a CP antenna (purple). (b) CDF of tag SNR for \name\ (blue), a vertical (red), and a horizontal (purple) antenna.}}
\end{figure}

This microbenchmark evaluates \name's ability to measure an orientation-independent phase of an RFID tag.

\noindent
\textbf{Baseline:} We compared our design to the performance of an approach that uses typical circularly polarized antennas. We used an MTI MT-242025 CP patch antenna~\cite{PatchAntenna} and placed it at the same distance from the tags as our system.

\noindent
\textbf{Experiment:} We placed the reader and our tag 0.5~m apart, and rotated the tag's roll in intervals of 10\textdegree. At each angle, we measured the change in phase of the tag (relative to an initial orientation), and  repeated this process for multiple frequencies (since \name\ collects wideband channel estimates). We used a separate circularly polarized antenna for powering to ensure the tags were always powered. 

\noindent
\textbf{Result:}
Fig. \ref{fig:mb_phase} plots the CDF of the change in phase across all measurements for \name\  (blue) and the baseline (purple). The result shows that \name\ reads a consistent tag phase independent of orientation, with a median change in phase of 0.1 radians and a 90\textsuperscript{th} percentile of 0.2 radians\footnote{\textred{We also measured the phase stability over 10 minutes in a controlled environment. Our results showed that the phase did not drift during this period and that the standard deviation was 0.02 radians across all measurements.}}. In contrast, a circularly polarized antenna has a change in phase ranging from 0.03 to 3.09 radians. This is expected since the tag was rotated at most 90\textdegree, which would result in an expected change in phase of $\pi$ radians.\footnotemark\ This demonstrates the need for \name's approach to estimate an orientation-independent tag phase for use in localization.

\section{SNR vs Orientation}

Recall that \name\ leverages a CCP design to generate circularly polarized signals that can read tags across orientations, so we evaluated its ability to read tags across angles.

\footnotetext{The expected change is twice the tag's rotation because the transmitter and receiver are both RHCP, so the phase offset is experienced twice.}

\noindent
\textbf{Baselines:} We compared to two baselines: \emph{Vertical Linear} where a vertical linearly polarized signal was used to read and \emph{Horizontal Linear} where a horizontal signal was used. 

\noindent
\textbf{Experiment:} We placed the reader and tag 0.5m apart. We rotated the tag\textred{'s roll} from -180\textdegree\ to 180\textdegree\ in intervals of 10\textdegree. At each angle, we measured the SNR of the response. If the tag's CRC checksum was incorrect, then the tag could not be read and the measurement was assigned an SNR of -30dB.

\noindent
\textbf{Result:}
Fig.~\ref{fig:mb_snr} plots a CDF of the SNRs across all measurements for \name\ (blue), \emph{Vertical Linear} (red), and \emph{Horizontal Linear} (purple). We observe that \name\ achieves a 10\textsuperscript{th}, 50\textsuperscript{th}, and 90\textsuperscript{th} percentile SNR of 35dB, 37dB, and 41dB, respectively, showing that \name's approach performs consis- tently across all tag orientations. In contrast, \emph{Vertical Linear} and \emph{Horizontal Linear} both had a 30\textsuperscript{th} percentile of -30dB, i.e., they cannot read tags at more than 30\% of the angles due to polarization mismatch. This matches our earlier investigation in~\xref{sec:problem} where linearly polarized antennas were unable to read tags across 30\% of angles.

\section{Impact of LOS on Range}\label{sec:app_nlos}

\textred{We compared the localization accuracy of \name\ in LOS and NLOS conditions. }

\noindent
\textbf{Experiment:}
\textred{We used the same setup as described in~\xref{sec:nulling_range}. In the \emph{LOS} experiment, all tags were in the line of sight of the antennas. In the \emph{NLOS} experiment, all tags were placed behind cardboard boxes, blocking the line of sight from the antenna to the tags. Since all measurements are constrained to a single plane in this micro-benchmark, we focused on 2D localization accuracy and computed the error as the L2-norm between the ground truth and the estimated tag location.}

\begin{figure}[t]
    \centering
    \includegraphics[width=0.8\linewidth]{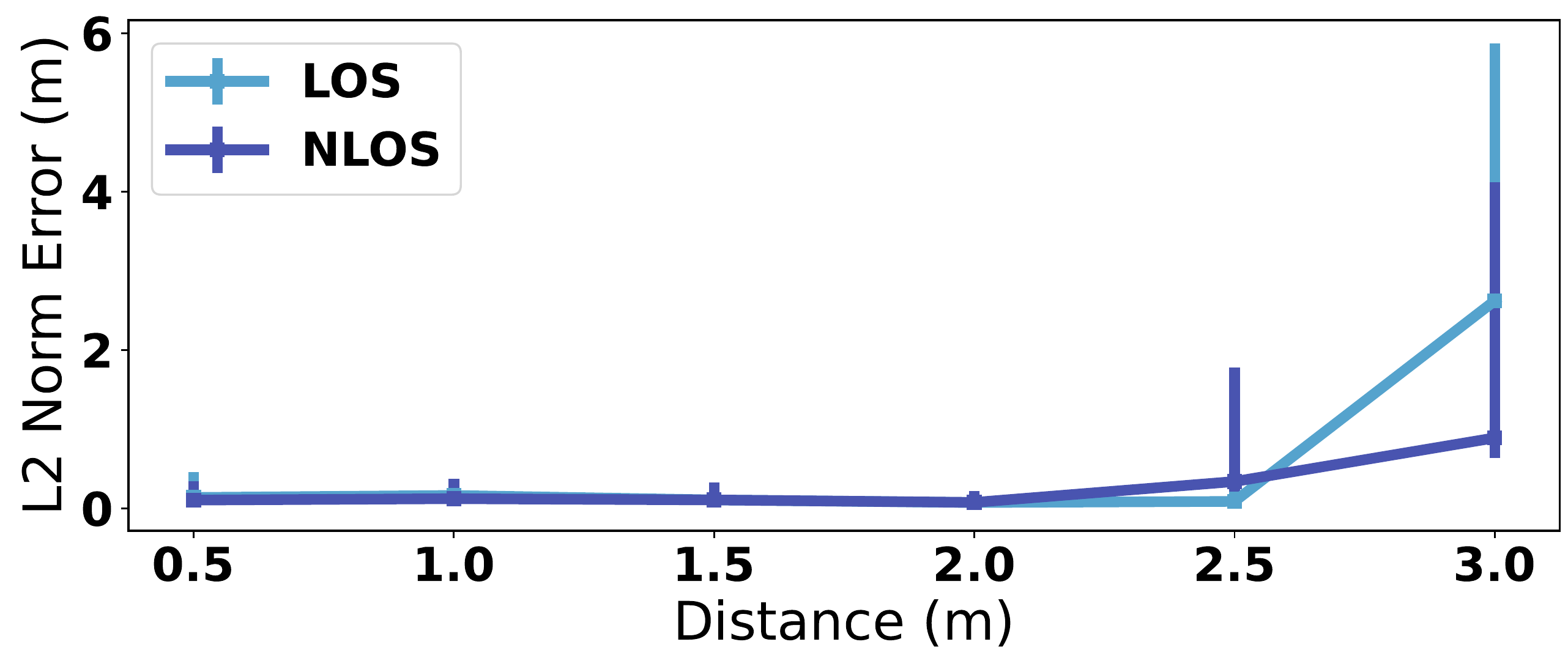}
    \caption{\textred{\footnotesize{Impact of LOS.} \textnormal{Median 2D localization error vs. distance in LOS(blue) and NLOS(purple).}}}
    \label{fig:nlos_distance}
\end{figure}

\noindent
\textbf{Result:}
\textred{Fig.~\ref{fig:nlos_distance} plots the median 2D error as a function of distance for \name\ in LOS (blue) and NLOS (purple) conditions. The error bars denote the 10\textsuperscript{th} and 90\textsuperscript{th} percentiles. We make the following remarks:}

\begin{itemize}
    \item For ranges up to 2~m, there is no significant difference between LOS and NLOS. For example, at 2~m, the median error is 7.6~cm for LOS and 10.8~cm for NLOS. This is expected since the SNR is high even for NLOS conditions.
    \item As the range increases to 2.5~m, the accuracy in NLOS conditions begins to degrade slightly, with a median error of 34~cm, compared to a median of 9~cm in LOS. This shows that \name\ performs similarly, but achieves a slightly shorter range in NLOS conditions compared to LOS conditions.  
    \item Surprisingly, at 3m, the accuracy in NLOS (0.89~m median) is better than in LOS (2.62~m median). This is due to the fact that low SNR at this range causes localization to fail in both LOS and NLOS, and the error in this case is highly variable (as shown by the large error bars at 3m). 
\end{itemize}

\section{Impact of VIO Tracking.}\label{sec:app_vio}

\begin{figure}[t]
    \centering
    \includegraphics[width=0.8\linewidth]{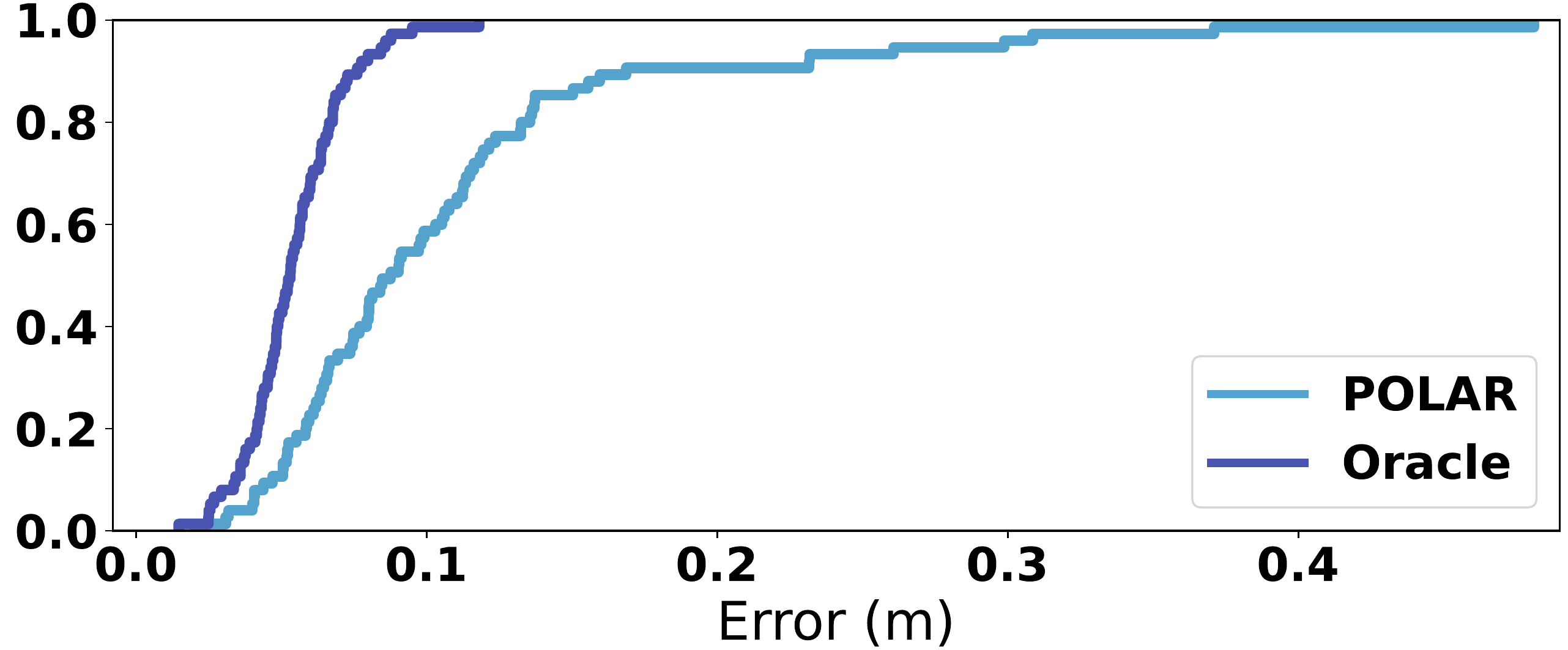}
    \caption{\textred{\footnotesize{Impact of VIO.} \textnormal{CDF of L2 norm error for \name(blue) and \emph{Oracle}(purple).  }}}
    \label{fig:vio_ablation}
\end{figure}

\textred{Recall from \xref{sec:implementation} that \name\ relies on VIO to track the location of the handheld reader over time. To evaluate the impact of potential VIO errors on the overall localization accuracy, we compared \name\ to an oracle system that uses the ground truth antenna locations when computing tag locations. We obtain ground truth antenna locations using the Optitrack\cite{optitrack} motion capture system. To ensure a fair comparison, we use the same measurements for the two systems, and only replace the VIO-derived locations with Optitrack-measured locations in Alg.~\ref{alg:selection}. }

\textred{Fig.~\ref{fig:vio_ablation} plots the CDF of the localization error for \name\ (blue) and the oracle (purple).}
\textred{We note that the oracle system achieves a 2~cm improvement in the median (5~cm vs 9~cm) and 9~cm improvement in the 90\textsuperscript{th} percentile (8~cm vs 17~cm). This is expected, since errors in the VIO self-tracking can contribute to \name's (already small) RFID localization error.}

\textred{It would be an interesting future direction to investigate other possible methods for device localization beyond VIO tracking. For example, perhaps fusing both VIO and self-localization with reference RFID tags (tags placed at known locations in the environment) could correct for VIO drift and enable even higher RFID localization accuracy. }

\end{document}